\documentclass[lettersize,journal]{IEEEtran}
\usepackage{amsmath,amsfonts,amssymb}
\usepackage{array}
\usepackage[caption=false,font=normalsize,labelfont=sf,textfont=sf]{subfig}
\usepackage{textcomp}
\usepackage{stfloats}
\usepackage{url}
\usepackage{verbatim}
\usepackage{graphicx}
\def\BibTeX{{\rm B\kern-.05em{\sc i\kern-.025em b}\kern-.08em
    T\kern-.1667em\lower.7ex\hbox{E}\kern-.125emX}}
\usepackage{balance}

\usepackage{amsthm}
\usepackage{xcolor}
\usepackage[style=ieee, backend=biber,maxnames=99]{biblatex}
\usepackage[colorlinks=true,
            linkcolor=blue,
            citecolor=blue,
            urlcolor=blue]{hyperref}

\usepackage{setspace}
\usepackage{multirow, hhline}

\usepackage[greek,english]{babel}
\usepackage[utf8]{inputenc}
\usepackage{bbm, dsfont}
\usepackage[linesnumbered,ruled,noline, noend]{algorithm2e}

\usepackage{float}

\theoremstyle{thmstyleone}%

\theoremstyle{thmstyletwo}%
\theoremstyle{thmstylethree}%
\newcommand{\multiRS}{\operatorname{multi-RS}}

\newcommand{\oners}{\operatorname{rs}}
\newcommand{\oneRS}{\operatorname{RS}}
\newcommand{\bconv}{\operatorname{BConv}}

\newcommand{\intt}{\operatorname{INTT}}
\newcommand{\ntt}{\operatorname{NTT}}
\newcommand{\iNtt}{\operatorname{(I)NTT}}

\newcommand{\relin}{\operatorname{Relin}}

\SetKwProg{Fn}{Procedure}{}{} % usage: \Fn{name(args)}{ ... }

\begin{document}
\title{Multi-Input Ciphertext Multiplication for Homomorphic Encryption}
\author{
Sajjad Akherati, and Xinmiao Zhang\\
		\IEEEauthorblockA{
The Ohio State University, Columbus, OH 43210, U.S.
\\
Emails: \{akherati.1, zhang.8952\}@osu.edu
}}
% \thanks{Manuscript created October, 2025; This work was developed by the IEEE Publication Technology Department. This work is distributed under the \LaTeX \ Project Public License (LPPL) ( http://www.latex-project.org/ ) version 1.3. A copy of the LPPL, version 1.3, is included in the base \LaTeX \ documentation of all distributions of \LaTeX \ released 2003/12/01 or later. The opinions expressed here are entirely that of the author. No warranty is expressed or implied. User assumes all risk.}
% }

\markboth{IEEE TRANSACTIONS ON CIRCUITS AND SYSTEMS—I: REGULAR PAPERS ,~Vol.~X, No.~X, August~2025}
{Multi-Input Ciphertext Multiplication for Homomorphic Encryption}

\maketitle

\begin{abstract}
Homomorphic encryption (HE) enables arithmetic operations to be performed directly on encrypted data. It is essential for privacy-preserving applications such as machine learning, medical diagnosis, and financial data analysis. In popular HE schemes, ciphertext multiplication is only defined for two inputs. However, the multiplication of multiple inputs is needed in many HE applications. In our previous work, a three-input ciphertext multiplication method for the CKKS HE scheme was developed. This paper first reformulates the three-input ciphertext multiplication to enable the combination of computations in order to further reduce the complexity. The second contribution is extending the multiplication to multiple inputs without compromising the noise overhead. Additional evaluation keys are introduced to achieve relinearization of polynomial multiplication results. To minimize the complexity of the large number of rescaling units in the multiplier, a theoretical analysis is developed to relocate the rescaling, and a multi-level rescaling approach is proposed to implement combined rescaling with complexity similar to that of a single rescaling unit. Guidelines and examples are provided on the input partition to enable the combination of more rescaling. Additionally, efficient hardware architectures are designed to implement our proposed multipliers. The improved three-input ciphertext multiplier reduces the logic area and latency by 15\% and 50\%, respectively, compared to the best prior design. For multipliers with more inputs, ranging from 4 to 12, the architectural analysis reveals 32\% savings in area and 45\% shorter latency, on average, compared to prior work. 
\end{abstract}


\vspace{-8pt}

\section{Introduction}\label{sec: introduction}
Homomorphic encryption (HE) enables computations to be performed directly on encrypted data without decrypting the ciphertexts. It is essential for privacy-preserving applications, such as machine learning, medical diagnosis, and financial data analysis, in the cloud. Popular HE schemes, including CKKS \cite{CKKS, RNSCKKS}, B/FV \cite{BV, FV}, and BGV \cite{BGV}, are based on the Ring Learning with Errors (RLWE) problem. A ciphertext consists of two polynomials, i.e., $\underline{\mathbf{ct}} = (\mathbf{c}_0(x), \mathbf{c}_1(x))$, each residing in the ring $\mathcal{R}_Q = \mathbb{Z}_Q[x]/(x^N + 1)$. In this ring, each polynomial has a degree of at most $N-1$, and the product of two polynomials is reduced modulo $x^N + 1$. The polynomial coefficients are reduced modulo $Q$. The parameter $N$ is typically a power of two on the order of thousands, and $Q$ is a modulus with a bit length of several hundreds to achieve a sufficient security level.

Long polynomial multiplication has been widely studied, and its complexity can be reduced by decomposing the polynomials and incorporating the modular reduction by $x^N+1$ into the decomposed components \cite{PolyMultSiPS, PolyMultJourn}. Alternatively, the Number Theoretic Transform (NTT) \cite{ParhiNTT, HanhoNTT, Area-efficient-conflict-free, AC-PM-poly-mult, area-efficient-ntt-HE} converts polynomial multiplications to individual coefficient multiplications. Integer modular multiplication can be simplified using Montgomery multiplication, Barrett reduction, and Karatsuba decomposition \cite{ZhangMultSiPS, MMIDSajjadZhangJSPS}. The residue number system (RNS) further reduces the complexity of integer arithmetic by decomposing an integer $a\mod Q$ into its residues modulo the co-prime factors of $Q$ \cite{RNSCKKS, FV-RNS}. 

The CKKS scheme \cite{CKKS} enables approximate arithmetic over real numbers and has demonstrated notable performance improvements in practical applications. The ciphertext multiplication of the CKKS scheme consists of three steps: polynomial multiplication, relinearization, and rescaling. 
Algorithmic reformulations have been developed to combine computations and thereby reduce the number of coefficient multiplications needed for the relinearization in the RNS-CKKS scheme \cite{Combined, Combined2}. Moreover, several hardware accelerators have been developed in prior works \cite{F1, ARK, BTS, SHARP}. Despite these efforts, ciphertext multiplication remains limited to two inputs. However, many applications require the multiplication of more than two ciphertexts, such as high-order polynomial approximations of the ReLU function \cite{Minimax_approx} in neural network inference \cite{HyPHEN}, decision trees for medical diagnosis \cite{medical-diagnosis-3}, and genome sequencing \cite{GenomeSequencing2}. Multiplying only two ciphertexts at a time results in a long latency. 

In the preliminary version of this work \cite{3-ct_mult}, a 3-input ciphertext multiplier for the RNS-CKKS scheme \cite{RNSCKKS} was developed. An additional evaluation key was introduced to relinearize the extra polynomial generated by multiplying the input ciphertext polynomials. The polynomial multiplication was reformulated using the Karatsuba method \cite{ZhangMult}, and the computations involved in the relinearization were combined to lower the overall complexity. An efficient, fully pipelined hardware architecture was also developed. Compared to multiplying three ciphertexts using two 2-input multipliers, it achieved substantial reductions in both area and latency. However, the design is limited to three input ciphertexts.

This paper makes two major contributions. First, for three-input multiplication in the RNS-CKKS scheme, the relinearization and rescaling are reformulated to combine intermediate computations, thereby reducing complexity. Second, the multiplication is extended to support $n>3$ inputs without compromising the noise overhead. Additional evaluation keys are introduced to relinearize the results of polynomial multiplications. The proposed $n$-input multiplier is implemented using a tree structure composed of smaller input multipliers. Rescaling must be performed after each multiplication in the tree to prevent exponential noise growth, which results in a significant number of rescaling units as $n$ increases. A theoretical analysis is developed to identify the conditions under which rescaling can be delayed until after the next multiplier and accordingly merged with the subsequent rescaling operation. Then, a multi-rescaling algorithm is proposed to implement combined rescaling. Further reformulations are introduced to reduce the number of required NTT and inverse (I-)NTT operations, such that the complexity of the combined $\mu$-rescaling remains similar to that of a single rescaling, regardless of the value of $\mu$. Guidelines and examples are provided for constructing the multiplier tree to combine as many rescaling operations as possible. Furthermore, efficient, fully-pipelined hardware implementation architectures are developed for our proposed designs. For RNS-CKKS with $N=2^{16}$ and $Q$ decomposed into 24 co-primes of 64 bits, the proposed three-input ciphertext multiplier reduces the logic area, memory requirements, and latency by 15\%, 20\%, and 50\%, respectively, compared to our preliminary work in \cite{3-ct_mult}. Moreover, for the number of input ciphertexts ranging from $4$ to $12$, the proposed method achieves 32\% lower area and 45\% shorter latency, on average, compared to applying the two-input multiplication \cite{RNSCKKS} in a binary tree.

The organization of this paper is as follows. Section \ref{sec: background} reviews background knowledge. Section \ref{sec: 3-ct mult} presents our preliminary 3-input ciphertext multiplier design. Section \ref{sec: low-complexity ciphertext multiplication} introduces our improved three-input ciphertext multiplier. The ciphertext multiplication algorithm is extended to multiple inputs in Section
\ref{sec: multi-input ciphertext multiplication algorithm}. This section also proposes a combined rescaling and provides guidelines for optimized multi-input multiplier design. Hardware complexity analyses and comparisons are provided in Section \ref{sec: complexity analyses and comparisons}. Section \ref{sec: conclusion} concludes the paper. 

\section{Preliminaries} \label{sec: background}
This section introduces the notations and reviews essential information of the RNS-CKKS scheme \cite{RNSCKKS, 3-ct_mult}, in particular, ciphertext multiplication. 

\begin{table}[t]
% \color{red}
\begin{center}
\caption{List of symbols used in this paper.}\label{tab: list of symbols}
\vspace{-15pt}
\setlength{\tabcolsep}{3pt}
\begin{tabular}{c c}
\hline
Symbol & Definition
\\ \hline
$\mathcal{R}_Q$ & Polynomial ring $\mathbb{Z}_Q(x)/(x^N+1)$
\\ 
$Q$ & Coefficient modulus of $\mathcal{R}_Q$
\\ 
$N$ & Polynomial degree of $\mathcal{R}_Q$
\\ 
$P$ & Modulus for key switching
\\ 
$L$ & Number of factors of $Q$
\\ 
$K$ & Number of factors of $P$
\\ 

$\mathbf{a}$ & A polynomial in the ring $\mathcal{R}_Q$
\\ 
$\mathbf{A}$ & NTT transformation of polynomial $\mathbf{a}$ 
\\ 
$\underline{\mathbf{ct}}$ & A ciphertext in $\mathcal{R}_Q^2$
\\ 
$\mathcal{U}_q$ & Uniform distribution in $[0, q)$
\\ 
$\mathcal{N}(0, \sigma^2)$ & Normal distribution with mean $0$ and variance
$\sigma^2$
\\ \hline
\end{tabular}
\vspace{-15pt}
\end{center}
\end{table}

 In this paper, polynomials in the ring are denoted by boldface letters (e.g., $\mathbf{c_0}$ and $\mathbf{c_1}$). A tuple of polynomials in the ring is denoted by a boldface letter with an underline (e.g., $\underline{\mathbf{ct}}=(\mathbf{c_0}, \mathbf{c_1})$). Polynomials transformed to the NTT domain are written in uppercase boldface letters (e.g. $\mathbf{A}=\ntt(\mathbf{a})$). The discrete uniform distribution in the range of $[0,q)$ and normal distribution with mean zero and variance $\sigma^2$ are denoted by $\mathcal{U}_q$ and $\mathcal{N}(0,\sigma^2)$, respectively. 

In the CKKS scheme, a ciphertext is represented as $\underline{\mathbf{ct}}=(\mathbf{c_0}, \mathbf{c_1})$, where $\mathbf{c_0}, \mathbf{c_1} \in \mathcal{R}_Q$. To reduce the complexity of modular reduction on the polynomial coefficients, the RNS representation decomposes the modulus $Q$ into the product of $q_0 q_1 \cdots q_{L-1}$, where $q_j$ ($0 \leq j < L$) are pairwise co-prime integers of similar bit length and are referred to as the RNS moduli. Accordingly, an integer $\alpha \pmod Q$ can be uniquely represented as $\left\{\alpha^{(j)} = \alpha \pmod {q_j} \mid 0 \leq j < L\right\}$. Consequently, a multiplication/addition on $\alpha$ can be performed as $L$ independent multiplications/additions on $\alpha^{(j)}$ \cite{Combined}. A list of symbols used in this paper has been provided in Table \ref{tab: list of symbols}.

In the CKKS scheme \cite{RNSCKKS}, the secret key is $\underline{\mathbf{sk}}=(1, \mathbf{s})$, where $\mathbf{s}$ is a random polynomial of degree $N-1$ with coefficients in $\{0,\pm1\}$ and Hamming weight $h$. The value of $h$ depends on the desired security level \cite{HE-AES}. The public key in the RNS representation is given by
\begin{align*} \label{eq: pk}
    \underline{\mathbf{pk}}=\left\{\underline{\mathbf{pk}}^{(j)}=\left(\mathbf b^{(j)},\mathbf a^{(j)}\right)\in\mathcal{R}_{q_j}^2\mid 0\leq j< L\right\}.
\end{align*}
The coefficients of $\mathbf a^{(j)}$ are drawn from $\mathcal{U}_{q_j}$, and $\mathbf b^{(j)}$ is computed as $-\mathbf a^{(j)}\mathbf s+\mathbf e$, where the coefficients of $\mathbf e$ are sampled from $\mathcal{N}(0,\sigma^2)$ and $\sigma$ is chosen according to the target security level \cite{HE-AES}. The plaintext polynomial $\mathbf m$ is first scaled by a positive integer $\Delta$, so that it is not affected by noise terms of low norm. To encrypt $\Delta \mathbf{m}$, a random polynomial $\mathbf{v}$ with coefficients in $\{0, \pm1\}$ and weight $N/2$ is first generated. In addition, the coefficients of $\mathbf e_0$ and $\mathbf e_1$ are drawn from $\mathcal{N}(0, \sigma^2)$. The ciphertext for $\mathbf m$ is then computed as
\begin{align*}
   \underline{\mathbf{ct}} = \left\{\underline{\mathbf c}^{(j)} = \left(\mathbf c_0^{(j)},\mathbf c_1^{(j)}\right)\in\mathcal{R}_{q_j}^2\mid 0\leq j< L\right\}, 
\end{align*}
where
\begin{align*}
    \left(\mathbf c_0^{(j)}, \mathbf c_1^{(j)}\right) = \left(\Delta\mathbf m+\mathbf v\mathbf b^{(j)}+\mathbf e_0, \mathbf v\mathbf a^{(j)}+\mathbf e_1\right).
\end{align*}
Given a ciphertext $\underline{\mathbf{ct}}$, its decryption is carried out as $\Delta\mathbf m\approx \mathbf c_0^{(0)}+\mathbf c_1^{(0)}\mathbf s$.

\begin{figure*}[t]
		\centering
		\includegraphics[scale=0.56]{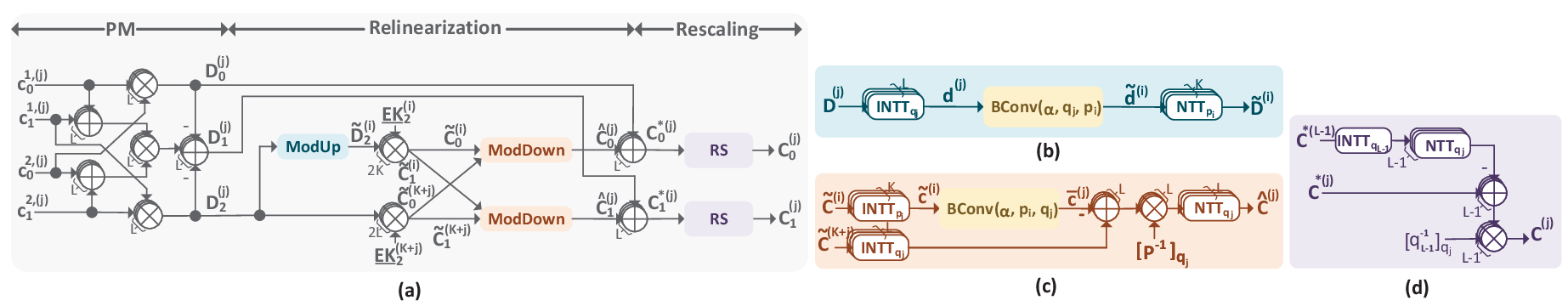}
        \vspace{-5pt}
        \caption{Hardware architectures for: (a) 2-input ciphertext multiplication using the RNS-CKKS scheme \cite{3-ct_mult}; (b) ModUp operation; (c) ModDown operation; and (d) rescaling (RS).}\label{fig: 2 ct mult arc}
        \vspace{-10pt}
\end{figure*}

The multiplication of two ciphertexts, $\underline{\mathbf{ct}}^1=\{\underline{\mathbf c}^{1,(j)}\mid 0\leq j<L\}$ and $\underline{\mathbf{ct}}^2=\{\underline{\mathbf c}^{2,(j)}\mid 0\leq j<L\}$, in the RNS-CKKS scheme \cite{RNSCKKS} consists of three major steps: polynomial multiplication (PM), relinearization, and rescaling (RS), as shown in Fig.~\ref{fig: 2 ct mult arc}(a). The PM step computes
\begin{equation}\label{eq: 2-pm karatsuba}
    \begin{aligned}
        \left(\!\mathbf d_0^{(j)}\!\!, \!\mathbf d_1^{(j)}\!\!, \!\mathbf d_2^{(j)}\!\right) \!\!=\!\! \left(\!\mathbf c_0^{1\!,\!(j)}\!\mathbf c_0^{2\!,\!(j)}\!\!\!,\! \mathbf c_0^{1\!,\!(j)}\!\mathbf c_1^{2\!,\!(j)}\!\!\!+\!\!\mathbf c_1^{1,\!(j)}\!\mathbf c_0^{2,\!(j)}\!\!\!,\! \mathbf c_1^{1,(j)}\!\mathbf c_1^{2,\!(j)}\!\right)\!\!.
    \end{aligned}
\end{equation}
Utilizing the Karatsuba formula, the number of polynomial multiplications in the equation above is reduced from 4 to 3 as shown in Fig.~\ref{fig: 2 ct mult arc}(a). The ciphertext's dimension grows into three polynomials as shown in \eqref{eq: 2-pm karatsuba}. It is reduced back to 2-polynomial format through $(\mathbf c_0^{*(j)}, \mathbf c_1^{*(j)})=(\mathbf d_0^{(j)},\mathbf d_1^{(j)})+\relin(\mathbf d_2^{(j)})$, where $\relin(\cdot)$ denotes relinearization. It utilizes an evaluation key ($\underline{\mathbf{ek}}_2\in\mathcal{R}^2_{PQ}$) with modulus $PQ$, where $P$ has a bit length similar to that of $Q$. Assume that $P$ is decomposed into $K\geq L$ co-prime factors as $P=\prod_{i=0}^{K-1}p_i$. From the secret key $\mathbf s$, the RNS components of $\underline{\mathbf{ek}}_2$ are generated as
\begin{equation}\label{eq: evks of ct mult for s^2}
\begin{aligned}
    &\underline{\mathbf{ek}}^{(i)}_2 = \left(\mathbf{ek}_{2,0}^{(i)},\mathbf{ek}_{2,1}^{(i)}\right)=\left(-\mathbf{ek}_{2,1}^{(i)}\mathbf{s}+\mathbf e_2, \mathbf{ek}_{2,1}^{(i)}\right), \\
    &\underline{\mathbf{ek}}^{(\!K\!+\!j\!)}_2 \!\!=\!\! \left(\!\mathbf{ek}_{2,0}^{(\!K\!+\!j\!)}\!\!,\! \mathbf{ek}_{2,1}^{(\!K\!+\!j\!)}\!\right) \!\!=\!\! \left(\!\!P\mathbf s^2\!\!-\!\!\mathbf{ek}_{2,1}^{(\!K\!+\!j\!)}\mathbf{s}\!+\!\mathbf e_2, \mathbf{ek}_{2,1}^{(\!K\!+\!j\!)}\!\right)\!\!,
\end{aligned}
\end{equation}
where the coefficients of $\mathbf{ek}_{2,1}^{(i)}$ and $\mathbf{ek}_{2,1}^{(K+j)}$ are drawn from $\mathcal{U}_{p_i}$ and $\mathcal{U}_{q_j}$, respectively, and those of $\mathbf e_2$ are sampled from $\mathcal N(0,\sigma^2)$. Using the evaluation key, the relinearization of $\mathbf d_2$ in the standard domain is defined as 
\begin{align}\label{eq: relin formula}
    \relin(\mathbf d_2) \triangleq \left(\left\lfloor \frac{\mathbf d_2\mathbf{ek}_{2,0}}{P}\right\rceil, \left\lfloor \frac{\mathbf d_2\mathbf{ek}_{2,1}}{P}\right\rceil\right)\mod{Q},
\end{align}
where the polynomial multiplications are performed in $\mathcal{R}_{PQ}$. After the division by $P$ and rounding, the coefficients are reduced modulo $Q$. 

RNS is employed to reduce the complexity of coefficient multiplications. However, $\underline{\mathbf{ ek}}_2$ is defined over a larger modulus $PQ$. Hence, the modulus of $\mathbf d_2$, which is $Q$, must be raised to $PQ$ using the ModUp procedure \cite{RNSCKKS} before multiplication with $\underline{\mathbf{ek}}_2$.
On the other hand, the division by $P$ and the modular reduction by $Q$ are handled by the ModDown process \cite{RNSCKKS}. Both ModUp and ModDown are realized through a basis conversion procedure. To reduce the complexity, a fast basis conversion algorithm was proposed in \cite{FV-RNS}. In particular, to convert the basis of $\mathbf d_2$ from $Q$ to $P$, it computes

\begin{align}\label{eq: basis conversion}
   \tilde{\mathbf{d}}_2^{(i)}\!=\!\bconv(\mathbf d_2^{(j)}, q_j, p_i) \!=\! \left[\sum_{j=0}^{L-1}\left[\hat{q}_j^{-1}\mathbf {d}_2^{(j)}\right]_{q_j}\hat{q}_j\right]_{p_i},
\end{align}
with $[\cdot]_{q_j}$ denoting modulo-$q_j$ operation, and $\hat{q}_j = Q/q_j$. Specifically, $\tilde{\mathbf d}_2^{(i)}$ denote the RNS components with respect to $p_i$. Together with the RNS components of $q_j$, they form the RNS representation with respect to modulus $PQ$. After the multiplication by $\underline{\mathbf{ek}}_2$, the resulting ciphertext $\underline{\tilde{\mathbf{ct}}}\in\mathcal{R}^2_{PQ}$ is divided by $P$ and reduced modulo $Q$ through the ModDown operation:
\begin{equation}\label{eq: moddown operation}
    \begin{aligned}
        &\hat{\mathbf c}_l^{(j)} = \left[P^{-1}\left(\tilde{\mathbf c}_l^{(K+j)}-\overline{\mathbf c}_l^{(j)}\right)\right]_{q_j}.
    \end{aligned}  
\end{equation}

Here, $\overline{\mathbf c}_l^{(j)} = \bconv(\tilde{\mathbf c}_l^{(i)}, p_i, q_j)$ ($l \in {0,1}$). The hardware architectures for the ModUp and ModDown operations are illustrated in Figs.~\ref{fig: 2 ct mult arc}(b) and~\ref{fig: 2 ct mult arc}(c), respectively. Since BConv is a nonlinear operation, INTT and NTT are required before and after BConv, respectively, when NTT is employed to simplify polynomial multiplications.

The noise added to ciphertexts increases after each multiplication. In the CKKS scheme, a rescaling operation is applied afterward to reduce the noise level, as follows
\begin{align}\label{eq: 1-rs}
    \oners(\mathbf c_l^{*(j)}) = \left[q_{L-1}^{-1}\left(\mathbf c_l^{*(j)}-\mathbf c_l^{*(L-1)}\right)\right]_{q_j}, (0\!\leq\! j\!<\!L\!\!-\!\!1).
\end{align}
In \eqref{eq: 1-rs}, the modulus of $c_l^{(L-1)}$ is $q_{L-1}$, which differs from $q_j$ for $0 \le j < L-1$. Furthermore, the ciphertext multiplication result must remain in the NTT domain for succeeding computations. Taking into account the NTT, the rescaling is performed as
\begin{align} \label{eq: 1 rs NTT}
    \oneRS(\!\mathbf c_l^{*(j)}\!)\!\! = \!\! \left[\!q_{L\!-\!1}^{-1}\!\!\left(\!\mathbf C_l^{*(j)}\!\!-\!\!\ntt_{q_j}\!\!\!\left(\!\intt_{q_{L\!-\!1}}\!\!\!\left(\!\mathbf C_l^{*(L\!-\!1)}\!\right)\!\!\right)\!\!\right)\!\!\right]_{q_j}\!\!\!\!,
\end{align}
where $\iNtt_{q_j}(\cdot)$ denotes (I)NTT with respect to $q_j$. The hardware architecture for RS unit is shown in Fig.~\ref{fig: 2 ct mult arc}(d). 

\begin{figure*}[t]
		\centering
		\includegraphics[scale=0.56]{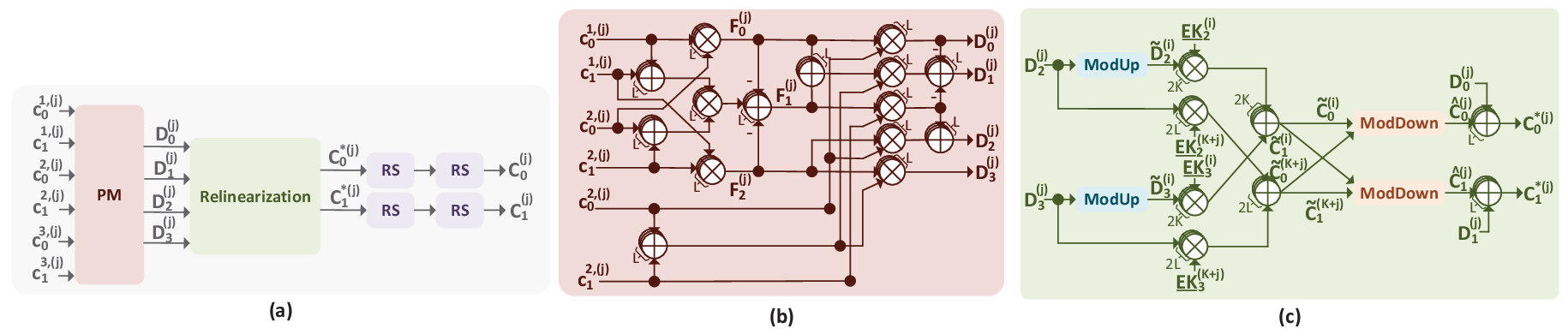}
        \vspace{-5pt}
        \caption{(a) Block diagram for three-input ciphertext multiplication using the RNS-CKKS scheme \cite{3-ct_mult}; (b) architecture for polynomial multiplication (PM) in three-input ciphertext multiplication; (c) architecture for relinearization. The ModUp, ModDown, and RS  blocks are implemented using the architecture in Fig. \ref{fig: 2 ct mult arc}(b), (c), and (d), respectively.} \label{fig: 3 ct mult arc}
        \vspace{-10pt}
\end{figure*}

\section{Preliminary Three-input Ciphertext Multiplication} \label{sec: 3-ct mult}
In our preliminary work \cite{3-ct_mult}, a three-input ciphertext multiplier for the RNS-CKKS scheme was developed. Denote the three input ciphertexts by $\underline{\mathbf{ct}}^1$, $\underline{\mathbf{ct}}^2$, and $\underline{\mathbf{ct}}^3$. They are decrypted as $\mathbf{c}^1_0+\mathbf{c}^1_1\mathbf{s}$, $\mathbf{c}^2_0+\mathbf{c}^2_1\mathbf{s}$, and $\mathbf{c}^3_0+\mathbf{c}^3_1\mathbf{s}$, respectively. The product of these three plaintext polynomials expands to $(\mathbf c^1_0\!+\!\mathbf c^1_1 \mathbf s)(\mathbf c^2_0\!+\!\mathbf c^2_1 \mathbf s)(\mathbf c^3_0\!+\!\mathbf c^3_1 \mathbf s)=\mathbf d_0\!+\!\mathbf d_1\mathbf s\!+\!\mathbf d_2 \mathbf s^2\!+\!\mathbf d_3 \mathbf s^3$, where 
\begin{equation}\label{eq: 3-ct polynomials}
\begin{aligned}
    &\mathbf d_0=\mathbf c_{0}^{1}\mathbf c_{0}^{2}\mathbf c_{0}^{3},\quad \mathbf d_1 = \mathbf c_{1}^{1}\mathbf c_{0}^{2}\mathbf c_{0}^{3} + \mathbf c_{0}^{1}\mathbf c_{1}^{2}\mathbf c_{0}^{3} + \mathbf c_{0}^{1}\mathbf c_{0}^{2}\mathbf c_{1}^{3},\\
    &\mathbf d_2 = \mathbf c_{1}^{1}\mathbf c_{1}^{2}\mathbf c_{0}^{3} + \mathbf c_{1}^{1}\mathbf c_{0}^{2}\mathbf c_{1}^{3} + \mathbf c_{0}^{1}\mathbf c_{1}^{2}\mathbf c_{1}^{3},\quad\mathbf d_3=\mathbf c_{1}^{1}\mathbf c_{1}^{2}\mathbf c_{1}^{3}.
\end{aligned}
\end{equation}
Apparently, $\mathbf d_0$ and $\mathbf d_1$ contribute to the $\mathbf c^*_0$ and $\mathbf c^*_1$ parts of the ciphertext product, respectively. Similar to that in 2-input ciphertext multiplication, $\mathbf d_2$ is relinearized by using the evaluation key $\underline{\mathbf{ek}}_2$ in \eqref{eq: evks of ct mult for s^2}. In this case, decrypting $\lceil P^{-1}\mathbf d_2 \underline{\mathbf{ek}}_2\rfloor$ leads to $\lceil P^{-1}\mathbf d_2(-\mathbf s\cdot \mathbf{ek}_{2,1}+\mathbf e+P\mathbf s^2+\mathbf s\cdot \mathbf{ek}_{2,1})\rfloor = \lceil P^{-1}\mathbf d_2(\mathbf e+P\mathbf s^2)\rfloor \approx \mathbf d_2\mathbf s^2$. In \cite{3-ct_mult}, it was proposed to introduce another evluation key $ \underline{\mathbf{ek}}_{3}$ to relinearize $\mathbf d_3$. In RNS format,
\begin{equation}\label{eq: evks of ct mult for s^3}
\begin{aligned}
    &\underline{\mathbf{ek}}^{(i)}_3 = \left(\mathbf{ek}_{3,0}^{(i)}, \mathbf{ek}_{3,1}^{(i)}\right)=\left(-\mathbf{ek}_{3,1}^{(i)}\mathbf{s}+\mathbf e_3, \mathbf{ek}_{3,1}^{(i)}\right),\\
    &\underline{\mathbf{ek}}^{(\!K\!+\!j\!)}_3 \!\!=\!\! \left(\!\mathbf{ek}_{3,0}^{(\!K\!+\!j\!)}\!\!\!, \mathbf{ek}_{3,1}^{(\!K\!+\!j\!)}\!\right) \!\!=\!\! \left(\!\!P\mathbf s^3\!\!-\!\!\mathbf{ek}_{3,1}^{(\!K\!+\!j\!)}\mathbf{s}\!+\!\mathbf e_3, \mathbf{ek}_{3,1}^{(\!K\!+\!j\!)}\!\right)\!\!,
\end{aligned}
\end{equation}
where the coefficients of $\mathbf{ek}_{3,1}^{(i)}$, $\mathbf{ek}_{3,1}^{(K+j)}$, and $\mathbf e_3$ follow the same distributions as the counterparts in $\underline{\mathbf{ek}}_2$ for $\mathbf d_2$ relinearization in \eqref{eq: evks of ct mult for s^2}. Accordingly, $\lceil P^{-1}\underline{\mathbf{ek}}_3\mathbf{d}_3\rfloor$ decrypts to $\mathbf{d}_3\mathbf{s}^3$. Overall, the final result of the three-input multiplier is 
\begin{equation}\label{eq: product3}
\underline{\mathbf{ct}}^*=(\mathbf{d}_0,\mathbf{d}_1)+ \lceil P^{-1}\mathbf{d}_2\underline{\mathbf{ek}}_2+P^{-1}\mathbf{d}_3\underline{\mathbf{ek}}_3\rfloor.
\end{equation}
RNS is adopted to reduce computation complexity, where the ModUp and ModDown operations are performed for modulus switching before and after the evaluation key multiplications, respectively.

In the original CKKS, rescaling is applied after every two-input ciphertext multiplication. In the case of the three-input multiplier, to achieve a similar noise level, two rescalings using the moduli $q_{L-1}$ and $q_{L-2}$ are applied to $\underline{\mathbf{ct}}^*$.

The architectures for implementing the three-input ciphertext multiplication in \cite{3-ct_mult} are shown in Fig. \ref {fig: 3 ct mult arc}(c). The PM block computes the products in \eqref{eq: 3-ct polynomials}. Utilizing the Karatsuba formula, the number of polynomial multiplications is reduced. Relinearizing the polynomials $\mathbf d_2^{(j)}$ and $\mathbf d_3^{(j)}$ requires four ModDown operations on $\mathbf d_2^{(j)}\mathbf{ek}_{2,0}^{(j)}$, $\mathbf d_2^{(j)}\mathbf{ek}_{2,1}^{(j)}$, $\mathbf d_3^{(j)}\mathbf{ek}_{3,0}^{(j)}$, and $\mathbf d_3^{(j)}\mathbf{ek}_{3,1}^{(j)}$. However, the term $\lceil P^{-1}\mathbf{d}_2\underline{\mathbf{ek}}_2 + P^{-1}\mathbf{d}_3\underline{\mathbf{ek}}_3 \rfloor$ in \eqref{eq: product3} can be reformulated as
$\lceil P^{-1}(\mathbf d_2 \mathbf{ek}_{2,0} + \mathbf d_3 \mathbf{ek}_{3,0}, \mathbf d_2 \mathbf{ek}_{2,1} + \mathbf d_3 \mathbf{ek}_{3,1})\rfloor$.
In this case, only two ModDown operations are needed for the two summations for each RNS component. This simplification is adopted to develop the relinearization architecture in Fig. \ref{fig: 3 ct mult arc}(c).

\section{Improved Three-Input Ciphertext Multiplication} \label{sec: low-complexity ciphertext multiplication}

\begin{figure*}[t]
		\centering
		\includegraphics[scale=0.56]{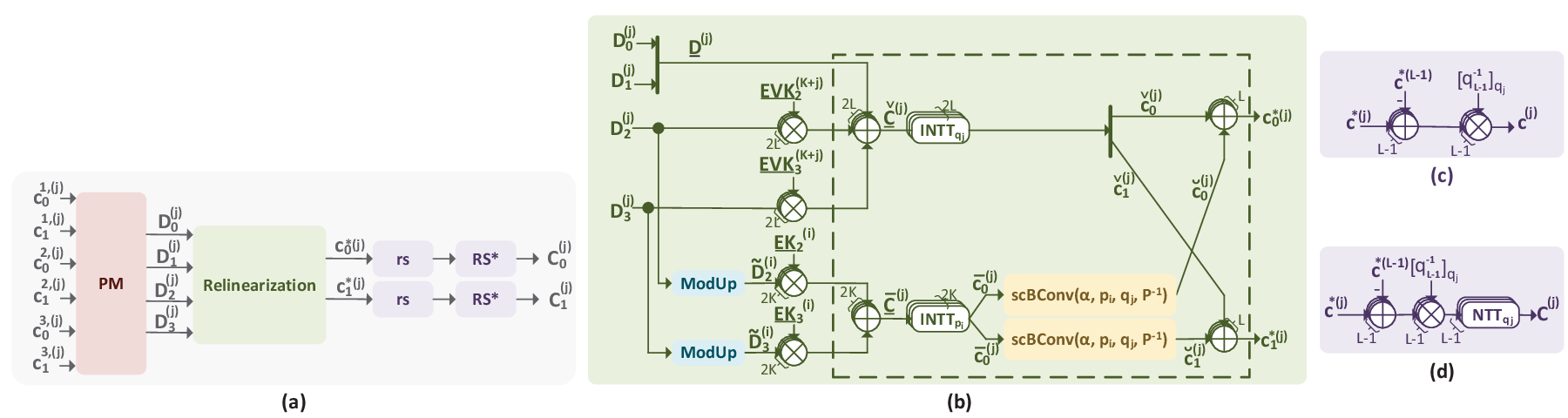}
        \vspace{-5pt}
        \caption{(a) Block diagram for the proposed improved three-input ciphertext multiplication; (b) architecture for relinearization; (c) architecture for rs block according to \eqref{eq: 1-rs}; and (d) the architecture for RS* block. The computations inside the dashed block in (b) correspond to those for the ModDown operation after applying the proposed improvements in Section \ref{sec: low-complexity ciphertext multiplication}}.\label{fig: 3 ct mult arc proposed low comp}
        \vspace{-10pt}
\end{figure*}

This section reformulates the computations involved in relinearization and rescaling for the three-input ciphertext multiplication. The overall architecture of the proposed design is shown in Fig. \ref{fig: 3 ct mult arc proposed low comp}. The PM block is identical to that in Fig. \ref{fig: 3 ct mult arc}(b), while the relinearization and rescaling blocks are simplified. Our new design reduces the number of (I)NTTs and shortens the data path by nearly half, as detailed below. 

Comparing \eqref{eq: 1-rs} and \eqref{eq: 1 rs NTT}, it can be observed that additional NTT and INTT operations are introduced into the rescaling process when the RNS form is adopted. As shown in Fig.~\ref{fig: 3 ct mult arc}(a) and (c), during the three-input ciphertext multiplication, the inputs to the two RS blocks connected to the relinearization block are given by $\mathbf{C}_l^{*(j)} = \mathbf{D}_l^{(j)} + \hat{\mathbf{C}}_l^{(j)}$ ($l \in {0,1}$). Tracing back to the ModDown architecture in Fig. \ref{fig: 2 ct mult arc}, $\hat{\mathbf{C}}_l^{(j)}$ are outputs of NTT blocks. By eliminating these NTTs in ModDown and adding INTT blocks to $\mathbf{D}_l^{(j)}$, which are generated by the polynomial multiplication, $\mathbf{c}_l^{*(j)}=\mathbf{d}_l^{(j)} + \hat{\mathbf{c}}_l^{(j)}$ is computed. Although this modification itself does not change the number of (I)NTTs needed for computing $\mathbf{C}_l^{*(j)}$ ($\mathbf{c}_l^{*(j)}$), now the rescaling can be carried out in the original domain according to \eqref{eq: 1-rs} by the rs block shown in Fig. \ref{fig: 3 ct mult arc proposed low comp}(c). Compared to the RS architecture in Fig. \ref{fig: 2 ct mult arc}(d), $L$ (I)NTT blocks have been eliminated in each rs block. The last rescaling block
(RS$^*$), whose architecture is shown in Fig. \ref{fig: 3 ct mult arc proposed low comp}(d), has NTT padded to the end so that the overall result is in NTT domain. Hence, it requires a similar number of (I)NTTs as the RS block.

Further simplifications can be made on the relinearization implementation. As mentioned in the previous paragraph, $D_l^{(j)}$ are passed through INTTs to get $d_l^{(j)}$, which are added to the output of the ModDown block in Fig. \ref{fig: 2 ct mult arc}(c). In this subfigure, $\tilde{C}^{(K+j)}_l$ also passes through INTTs. Intuitively, additions may be moved before the INTTs to reduce the total number of INTT blocks needed. To achieve this, the multiplication by $P^{-1}$ in Fig. \ref{fig: 2 ct mult arc}(c) is moved to before the adder.  Accordingly,
\begin{equation}\label{eq: moddown improved 1}
\begin{aligned}
\intt{q_j}(\mathbf{D}_l^{(j)}) + P^{-1}\intt{q_j}(\Tilde{\mathbf{C}}^{(K+j)}_l)
\end{aligned}
\end{equation}
can be computed prior to being added to $P^{-1}\Bar{\mathbf{c}}^{(j)}_l$ to obtain $\mathbf{c}^{*(j)}_l$.  Since $P^{-1}$ is a constant and INTT is linear, \eqref{eq: moddown improved 1} becomes 
\begin{align*}\label{eq: moddown improved 2}
\intt_{q_j}(\mathbf D_l^{(j)} + P^{-1}\Tilde{\mathbf{C}}^{(K+j)}_l).
\end{align*}
As a result, INTT no longer needs to be applied separately to $D_l^{(j)}$ and $\tilde C_l^{(K+j)}$. Since INTT is only applied to their sum, the number of INTT blocks is further reduced by $2L$.

Instead of multiplying $P^{-1}$ to each coefficient of $\Tilde{\mathbf{C}}^{(K+j)}_l$, it can be incorporated into the evaluation key and pre-multiplied. Similarly, the scaler multiplication of $P^{-1}$ can be also incorporated into basis conversion as 
\begin{equation}\label{eq: scBconv}
\begin{aligned}
&P^{-1}\bconv\left(\intt_{p_i}\left(\tilde{\mathbf C}_l^{(i)}\right),p_i,q_j\right) \\
    &\quad= \left[\sum_{i}^{K-1}\left[\hat{p}_i^{-1}\cdot \intt_{p_i}\left(\tilde{\mathbf{C}}_l^{(i)}\right)\right]_{p_i}{p}_i^{-1}\right]_{q_j}.
\end{aligned}
\end{equation}
These scaler pre-multiplications eliminate $2L$ constant multiplications.

The proposed modifications also shorten the data path. The architectures in Fig. \ref{fig: 3 ct mult arc} originally include eight (I)NTTs in the data path. Using our proposed modifications, the NTT in ModDown is eliminated since $\hat{c}_l^{(j)}$ instead of $\hat{C}_l^{(j)}$ is computed. Comparing with the RS block in Fig. \ref{fig: 2 ct mult arc}(d), the rs and RS$^*$ blocks in Fig. \ref{fig: 3 ct mult arc proposed low comp} (c) and (d), respectively, reduce the data path by two and one (I)NTTs, respectively. Overall, our improved design eliminates four (I)NTT operations from the data path.

%The architecture for the improved three-input ciphertext multiplication is shown in Fig.~\ref{fig: 3 ct mult arc proposed low comp}(a). It is composed of three main components: the PM block, the relinearization block, and the rescaling block. The PM block is identical to that in Fig.~\ref{fig: 3 ct mult arc}(b). The relinearization shown in Fig. \ref{fig: 3 ct mult arc proposed low comp}(b) is modified to reduce the number of INTTs and $P^{-1}$ multipliers \textcolor{red}{by taking into account the simplifications in \eqref{eq: moddown improved 2} and \eqref{eq: scBconv}}. The first rescaling (rs) is done in the original domain by the architecture in Fig. \ref{fig: 3 ct mult arc proposed low comp}(c). The second rescaling block (RS$^*$) is the last rescaling block and hence has NTT at the end so that the following computations can continue in the transformed domain. \textcolor{red}{Accordingly, the number of (I)NTT operations associated with the rescaling operations is also reduced by $2L$.}

\section{Multi-Input Ciphertext Multiplication with Minimized Multiplicative Depth} \label{sec: multi-input ciphertext multiplication algorithm}

The HE evaluation of many applications, such as the activation function for neural networks, decision trees, and genome sequencing, requires the multiplication of more than three ciphertexts. This section extends ciphertext multiplication to more than three inputs. Multiplying $n$ ciphertexts results in $n+1$ polynomials, and the evaluation keys are generalized to bring the result back to the 2-polynomial format. HE introduces noise to the ciphertext, and this noise grows with each multiplication. Since the scaling factor $\Delta$ is multiplied by the plaintext message during the encryption, the noise is also scaled by $\Delta$ in a two-input ciphertext multiplication. The magnitude of $\Delta$ is similar to that of each co-prime factor of $Q$. Hence, rescaling restores the noise level to its original magnitude. For a given $Q=q_0q_1\cdots q_{L-1}$, at most $L-1$ levels of 2-input ciphertext multiplications, each followed by rescaling, can be carried out without corrupting the message. Therefore, the multiplicative depth of the computations must be minimized to reduce the frequency of activating the expensive bootstrapping process for refreshing the ciphertext. Multiplying $n$ ciphertexts needs a multiplicative depth of $\lceil \log_2 n\rceil$ using a binary tree of two-input multipliers. The proposed $n$-input ciphertext multiplier is designed to handle multiple inputs with lower complexity and the same multiplicative depth.

\subsection{Multi-input Ciphertext Multiplication} \label{subsec: extended relinearization}
Consider $n$ ciphertexts in the CKKS scheme, denoted by $\underline{\mathbf{ct}}^u  = (\mathbf{c}_0^{u},\mathbf{c}_1^u)$ $(1\leq u \leq n)$. These ciphertexts are decrypted to $\mathbf{c}_0^{u}+\mathbf{c}_1^{u}\mathbf{s} \pmod{\mathcal R_{Q}}$. The product of the plaintexts is
\begin{align}    &\prod_{u=1}^{n}\left(\mathbf{c}_0^{u}+\mathbf{c}_1^{u}\mathbf{s} \right) = \sum_{t=0}^{n}\mathbf{d}_t\mathbf{s}^t \notag, \text{where }
    \mathbf{d}_t = \sum_{\substack{l_1,\dots,l_n \in \{0,1\} \\ l_1 + \cdots + l_n = t}} 
    \prod_{u=1}^{n} \mathbf{c}_{l_u}^{u}. \label{eq: d_u formula for n-ct mult} 
\end{align}
Each of $\mathbf{d}_t$ with $t>1$ needs to be relinearized to the 2-polynomial format. Similarly, this can be achieved by forming an evaluation key $\underline{\mathbf {ek}}_t$ such that decrypting $\lceil P^{-1}\mathbf d_t \underline{\mathbf {ek}}_t\rfloor$ leads to $\mathbf d_t \mathbf s^t$. The evaluation key satisfying this condition in the RNS format is given by
\begin{equation}\label{eq: evks of ct mult for s^t}
\begin{aligned}
    &\underline{\mathbf{ek}}^{(i)}_t = \left(\mathbf{ek}_{t,0}^{(i)}, \mathbf{ek}_{t,1}^{(i)}\right)=\left(-\mathbf{ek}_{t,1}^{(i)}\mathbf{s}+\mathbf e_t, \mathbf{ek}_{t,1}^{(i)}\right),\\
    &\underline{\mathbf{ek}}^{(\!K\!+\!j\!)}_t \!\!=\!\! \left(\!\mathbf{ek}_{t,0}^{(\!K\!+\!j\!)}\!\!\!, \mathbf{ek}_{t,1}^{(\!K\!+\!j\!)}\!\right) \!\!=\!\! \left(\!\!P\mathbf s^t\!\!-\!\!\mathbf{ek}_{t,1}^{(\!K\!+\!j\!)}\mathbf{s}\!+\!\mathbf e_t,\! \mathbf{ek}_{t,1}^{(\!K\!+\!j\!)}\!\right)\!\!,
\end{aligned}
\end{equation}
where $\mathbf{ek}_{t,1}$ and $\mathbf{e}_t$ have random coefficients following $\mathcal{U}_{PQ}$ and $\mathcal{N}(0,\sigma^2)$ distributions, respectively. Using the evaluation keys, the relinearization for the $n$-input ciphertext multiplication is defined as
\begin{align} \label{eq: extended relin}
    \underline{\mathbf{ct}}^* = (\mathbf{c}_0^*,\mathbf{c}_1^*) = (\mathbf{d}_0,\mathbf{d}_1) + \sum_{t=2}^{n}\left\lceil P^{-1}\mathbf d_t \underline{\mathbf {ek}}_t\right\rfloor.
\end{align}

Multiplying the $n$ terms of ($\mathbf{c}_0^{u}+\mathbf{c}_1^{u}\mathbf{s}$)
entails $2^n$ polynomial multiplications. To reduce the number of polynomial multiplications, the fast filtering techniques \cite{minimal-filtering-alg} can be iteratively applied. Similar methods as proposed in previous sections can also be adopted to simplify the implementation of the extended relinearization of \eqref{eq: extended relin} in RNS representation. The resulting architecture closely resembles that in Fig.~\ref{fig: 3 ct mult arc proposed low comp}(b), containing $n-1$ copies of the ModUp blocks, multipliers, and adders on its left side.

When the $n$ input ciphertexts are multiplied altogether, the noise in the product is scaled by $\Delta^{n-1}$. Consequently, its multiplicative depth becomes $n-1$, rather than $\lceil \log_2 n\rceil$ as in designs that multiply $n$ ciphertexts using 2-input multipliers arranged in a binary tree structure. To maintain the same multiplicative depth, rescaling must be performed on intermediate results before the relinearization to scale down the noise in our proposed $n$-input multiplier. In this case, the inputs to the rescaling units are the outputs of the polynomial multiplications in the NTT domain, and the rescaling can not be carried out in the original domain. As a result, numerous (I)NTTs are required to implement the rescaling. In the following subsections, first, a multi-rescaling method is proposed to implement an arbitrary number of concatenated rescaling operations with the same number of (I)NTTs as that in the first rescaling. Then, analyses and examples are provided to determine when rescaling blocks can be moved adjacent to one another to take advantage of the combined rescaling design.

\subsection{Multi-Rescaling } \label{subsec: multi-rs}
This section proposes an efficient multi-rescaling (multi-RS) method that implements $\mu$ ($1<\mu<L$) concatenated
rescaling with the same number of (I)NTTs as in the first rescaling. 

If the formula in \eqref{eq: 1 rs NTT} is utilized to implement each rescaling, a large number of (I)NTT operations would be required. Alternatively, one may transform the input polynomial back to the INTT domain using $L$ INTT operations, utilize \eqref{eq: 1-rs} to carry out each rescaling, and then apply $L-\mu$ NTT operations to convert the results back to the NTT domain for subsequent computations. This method requires $2L-\mu$ (I)NTT operations overall. Because (I)NTTs are hardware-intensive, reformulations are developed in the next section to reduce their number in the case of multiple concatenated rescaling.

\begin{algorithm}[t]
  \caption{$\multiRS\left(\mathbf{A}^{(j)},\mu\right)$}  \label{alg: multi-RS}
  \KwIn{
      $\mathbf{A}^{(j)}$ for $0 \leq j < L$; \\
      \hspace*{3.1em} Rescaling depth $\mu$; \\
      \hspace*{3.1em} precomputed constants  $g_\eta^{\mu,t}$ for $0 \leq \eta < L-\mu$, \\
      \hspace*{3.1em} and $L-\mu \leq t \leq L-1$.
    }

  \BlankLine
  % Step 1
  \For{$u \gets L-\mu$ \KwTo $L-1$}{
    $\mathbf{a}^{(u), \{0\}} \gets \intt_{q_u}\!\left(\mathbf{A}^{(u)}\right)$\;
  }

  % Step 2
  \For{$u \gets 1$ \KwTo $\mu-1$}{
    \For{$t \gets L-\mu$ \KwTo $L-1-u$}{
      $\mathbf{a}^{(t), \{u\}} \gets \oners\!\left(\mathbf{a}^{(t), \{u-1\}}\right)$ using \eqref{eq: 1-rs}\;
    }
  }

  % Step 3
  \For{$\eta \gets 0$ \KwTo $L-\mu-1$}{
    $\mathbf b^{(\eta)} \gets \sum_{t=L-\mu}^{L-1} g_\eta^{\mu,t}\,\mathbf{a}^{(t), \{L-1-t\}} \pmod{q_\eta}$\;
    $\mathbf{B}^{(\eta)} \gets \ntt_{q_\eta}\!\left(\mathbf{b}^{(\eta)}\right)$\;
    $\mathbf{A}^{(\eta), \{\mu\}} \gets g_{\eta}^{\mu,L-1}\mathbf{A}^{(\eta)} - \mathbf B^{(\eta)} \pmod{q_\eta}$\;
  }
  
  \BlankLine
  \KwOut{
    $\mathbf{A}^{(\eta),\{\mu\}}$ for $0 \leq \eta < L-\mu$.
  }
\end{algorithm}

For a set of polynomials in RNS representation $\{\mathbf{a}^{(j)}\mid 0\leq j<L\}$, denote the result of applying $\mu$ consecutive rescaling by $\{\mathbf a^{(\eta),\{\mu\}}\mid 0 \leq \eta < L-\mu\}$. For $0\leq\eta < L-u\leq t<L$, define the pre-computed constants
\begin{align} \label{eq: g(u,t,eta)}
       g^{u,t}_\eta \triangleq q_{L-u}^{-1}q_{L-u+1}^{-1}\cdots q_t^{-1} \pmod{q_\eta}.
\end{align}
Next, it will be shown by induction that for each $1\leq \mu<L$
\begin{align}\label{eq: multi rs}
           \mathbf a^{(\eta),\{\mu\}} \!\!=\!\!\left(\! g^{\mu, L-1}_ \eta\mathbf{a}^{(\eta)} \!\!-\!\! \!\!\!\sum_{t=L-\mu}^{L-1}\!\!\!\!g^{\mu,t}_\eta\mathbf{a}^{(t),\{L-1-t\}}\!\! \right)\!\!\!\!\!\mod{q_\eta}.      
\end{align}
From \eqref{eq: g(u,t,eta)}, it can be derived that $g_\eta^{1,L-1} = q_{L-1}^{-1}\mod{q_\eta}$. For the base case $\mu=1$, \eqref{eq: multi rs} becomes $\mathbf{a}^{(\eta),\{1\}} = g_\eta^{1,L-1}\mathbf a^{(\eta)}-g_\eta^{1,L-1}\mathbf a^{(L-1),\{0\}}  \mod{q_\eta} = q_{L-1}^{-1}\mathbf{a}^{(\eta)}-q_{L-1}^{-1}\mathbf a^{(L-1)}\mod{q_\eta}$, which is equal to the result of applying a single rescaling block from \eqref{eq: 1-rs}. Accordingly, \eqref{eq: multi rs} holds for the base case $\mu=1$. Suppose it also holds for $\mu-1$. Using \eqref{eq: 1-rs}, it can be derived that
\begin{align*}
    \mathbf{a}^{(\eta),\{\mu\}} &= \oners(\mathbf{a}^{(\eta),\{\mu-1\}})\\
    &= q_{L-\mu}^{-1}\mathbf{a}^{(\eta),\{\mu-1\}} - q_{L-\mu}^{-1}\mathbf{a}^{(L-\mu),\{\mu-1\}} \pmod{q_\eta}. 
\end{align*}
By plugging in $\mathbf{a}^{(\eta),\{\mu-1\}}$ from \eqref{eq: multi rs} in the above equation, $\mathbf{a}^{(\eta),\{\mu\}}$ is rewritten as
\begin{align*}
    \mathbf{a}^{(\eta),\{\mu\}} &= \bigg(q_{L-\mu}^{-1}\big(g^{\mu-1, L-1}_ \eta\mathbf{a}^{(\eta)} \!\!- \!\!\!\sum_{t=L-\mu+1}^{L-1}\!\!\!\!g^{\mu-1,t}_\eta\mathbf{a}^{(t),\{L-1-t\}}\big) \\
    &\quad - q_{L-\mu}^{-1}\mathbf{a}^{(L-\mu),\{\mu-1\}}\bigg) \mod{q_\eta}.
\end{align*}
From \eqref{eq: g(u,t,eta)}, it follows that $q_{L-\mu}^{-1}g_\eta^{\mu-1,t}\pmod{q_\eta} = g_\eta^{\mu,t}$ and $q_{L-\mu}^{-1}\pmod{q_\eta} = g_\eta^{\mu,L-\mu}$. Consequently, from the above equation, it follows that
\begin{align*}
    \mathbf{a}^{(\eta),\{\mu\}} &= \bigg(g^{\mu, L-1}_ \eta\mathbf{a}^{(\eta)} \!\!- \!\!\!\sum_{t=L-\mu+1}^{L-1}\!\!\!\!g^{\mu,t}_\eta\mathbf{a}^{(t),\{L-1-t\}} \\
    &\quad - g_\eta^{\mu,L-\mu}\mathbf{a}^{(L-\mu),\{\mu-1\}}\bigg) \mod{q_\eta}
    % \\
    % &=g^{\mu, L-1}_ \eta\mathbf{a}^{(\eta)} \!\!- \!\!\!\sum_{t=L-\mu}^{L-1}\!\!\!\!g^{\mu,t}_\eta\mathbf{a}^{(t),\{L-1-t\}}\pmod{q_\eta}. 
\end{align*}
Incorporating the last term into the summation in the above formula yields \eqref{eq: multi rs}.

In \eqref{eq: multi rs}, the modulus of $\mathbf{a}^{(\eta)}$ is also $q_{\eta}$; hence, (I)NTTs are not required for these components during computation. Meanwhile, each $\mathbf{a}^{(t)}$ for $L-\mu \leq t < L$ has a distinct modulus and must be derived through $\mu$ INTTs at the input stage. Then, $\mathbf{a}^{(t),{L-1-t}}$ are obtained by iteratively applying the rescaling operation in \eqref{eq: 1-rs}. For each $0 \leq \eta < L-\mu$, the sum of products in \eqref{eq: multi rs} is transformed back using one NTT and added to the scaled NTT of $\mathbf{a}^{(\eta)}$ at the end. Overall, only $L$ (I)NTT blocks are required. The proposed simplified multi-rescaling procedure is summarized in Algorithm~\ref{alg: multi-RS}.

\begin{figure}[t]
		\centering
		\includegraphics[scale=0.7]{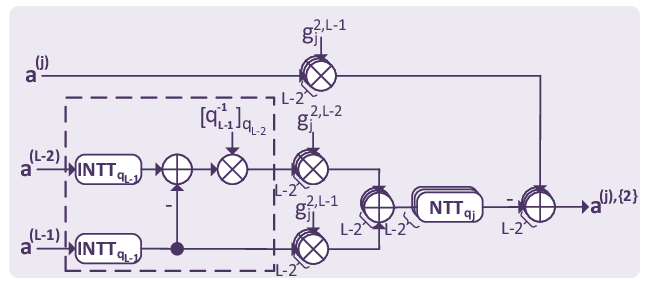}
        \vspace{-5pt}
        \caption{Hardware implementation architecture of the proposed multi-RS for $\mu=2$ (2-RS).}\label{fig: multi rs}
        \vspace{-10pt}
\end{figure}

Fig.~\ref{fig: multi rs} illustrates the hardware architecture of the proposed multi-rescaling algorithm for the example case of $\mu = 2$. The units within the dashed block implement Line~7 of Algorithm~\ref{alg: multi-RS}. The modulus of the multiplier in this block is $q_{L-2}$, and hence this multiplication cannot be merged with subsequent modular multiplications that use moduli $q_j$ for $0 \leq j < L-2$. The outputs of the dashed block are shared, whereas the other computation units in Fig.~\ref{fig: multi rs} are duplicated for the rescaling of each $\mathbf{a}^{(j)}$ with $0 \leq j < L-2$. Compared to the two concatenated rescaling blocks shown in Fig.~\ref{fig: 2 ct mult arc}(d), the total number of (I)NTTs is reduced from $2L-1$ to $L$, although the proposed design requires more integer modular multipliers. 

\vspace{-5pt}

\subsection{Input Partition and Rescaling Combination}

As mentioned previously, carrying out rescaling after multiplying all input polynomials in a multi-input ciphertext multiplication can increase the multiplicative depth. Hence, rescaling should be applied after every few polynomials are multiplied. This section provides guidelines and examples for the input partition to enable the adoption of additional multi-rescaling steps for minimizing complexity without increasing the multiplicative depth.

The multiplicative depth for an $n$-input ciphertext multiplication is $\lceil \log_2 n \rceil$. The $n$ inputs are partitioned into $m$ groups of sizes $n_1, n_2,\cdots,n_m$, where $n_1+n_2+\cdots+n_m=n$, and it is denoted as $(n_1,n_2,\cdots, n_m)$. Without loss of generality, assume that $n_1\geq n_2\geq \cdots \geq n_m$. This partitioning can be recursively applied across multiple layers. First, let us assume that the original single-level rescaling is used, which is applied after the polynomials in every group are multiplied. 

\begin{figure}[t]
		\centering
		\includegraphics[scale=0.58]{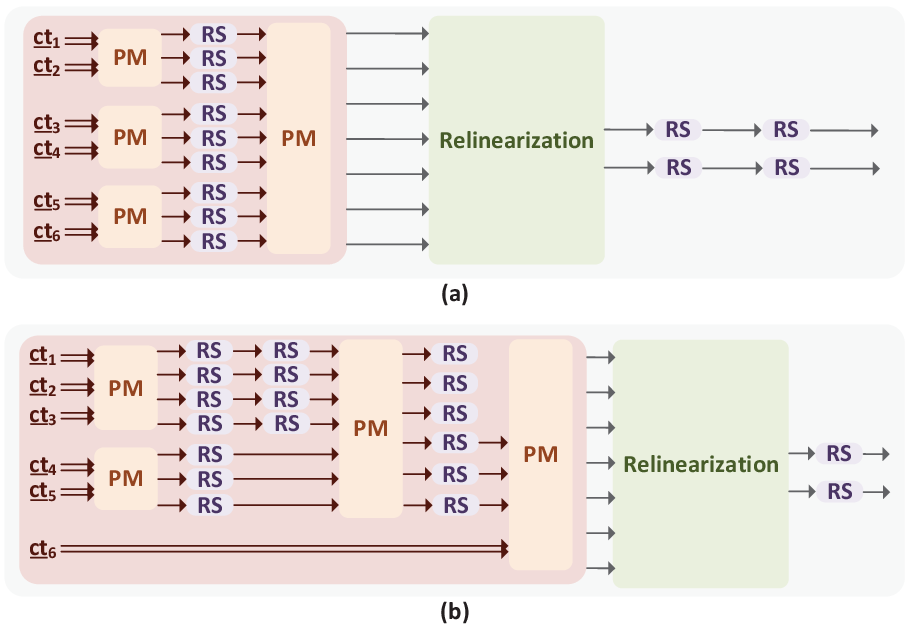}
        \vspace{-15pt}
        \caption{Block diagram for the 6-input ciphertext multiplication: (a) the ciphertexts are partitioned to $(2,2,2)$; (b) the ciphertexts are grouped as $(5,1)\mid(3,2)$.}
        \vspace{-15pt}
\label{fig: 6-CT mult}
\end{figure}

Consider an example of a six-input ciphertext multiplication. The multiplicative depth is $\lceil \log_2 6\rceil = 3$. This can be achieved using the partition shown in Fig.~\ref{fig: 6-CT mult}(a), where the six inputs are divided into groups of $(2,2,2)$. One rescaling block is required for each polynomial after every two-input multiplication to reduce the noise scalar from $\Delta^2$ to $\Delta$. Subsequently, two rescaling blocks are needed for each polynomial after multiplying the results of the three two-input groups to further reduce the noise from $\Delta^3$ to $\Delta$. To minimize the number of required rescaling units, the rescaling operation in the final multiplication stage can be deferred until after the relinearization, as shown in Fig.~\ref{fig: 6-CT mult}. Overall, the multiplicative depth remains three. Alternatively, the six inputs can be partitioned as $(5,1)$, where the group of five inputs is further divided into $(3,2)$. Fig.~\ref{fig: 6-CT mult}(b) illustrates the rescaling blocks required to bring the noise scalar back to $\Delta$ under this partitioning. After the multiplication of the three inputs, the noise scalar becomes $\Delta^3$, requiring two rescaling blocks for each polynomial to restore it to $\Delta$. Each of the other multiplications in Fig.~\ref{fig: 6-CT mult}(a) involves two operands, increasing the noise scalar to $\Delta^2$; hence, one rescaling block is needed for each polynomial after each multiplication. Overall, the data path from the three-input multiplication to the output contains four rescaling blocks, which is one level higher than $\lceil \log_2 6\rceil = 3$.

Consider the general partition of $n$ inputs into $m$ groups $(n_1,n_2,\cdots, n_m)$. Since the final multiplication of the $m$ group results scales the noise by $\Delta^{m-1}$, $m-1$ rescaling blocks are required at the end. Therefore, the multiplicative depth of each of the $m$ groups should not exceed $\lceil \log_2 n\rceil-(m-1)$. The partitioning can be further applied to the next layer. As the multiplicative depth is determined by the path with the largest number of rescalings, and $\max\{n_i\mid 1\leq i\leq m\}=n_1$, the multiplicative depth of each group should not exceed $\lceil \log_2 n_1\rceil-(m'-1)$, if that group is further divided into $m'$ subgroups. For instance, when the $6$ inputs are partitioned into groups of $(2,2,2)$ as in the previous example, $\lceil \log_2 n\rceil-(m-1)=\lceil \log_2 6\rceil-(3-1)=1$. Since $n_1=2$, its multiplication can be implemented by a depth of $\lceil \log_2 2\rceil=1$, which satisfies the condition. In contrast, when the partition is $(5,1)$, $\lceil \log_2 n\rceil-(m-1)=\lceil \log_2 6\rceil-(2-1)=2$. However, the multiplication of $n_1=5$ inputs needs a depth of at least $\lceil \log_2 5\rceil=3$, which violates the condition.

As shown in Section \ref{subsec: multi-rs}, our proposed multi-rescaling requires the same number of (I)NTTs as in a single rescaling, regardless of how many rescaling units are combined. Since (I)NTTs are much more complicated than the polynomial multiplications, the multi-rescaling operation exhibits a complexity similar to that of a single rescaling. Therefore, utilizing the multi-rescaling reduces the overall hardware complexity. Multi-rescaling becomes feasible when the inputs are partitioned into $m > 2$ groups, in which case there are $m-1$ rescaling units after multiplying the results of the $m$ groups, and they can be combined using our proposed design. For example, the two concatenated rescaling blocks in Fig. \ref{fig: 6-CT mult}(a) can be merged into a single two-rescaling (2-RS) unit.

\begin{figure}[t]
		\centering
		\includegraphics[scale=0.7]{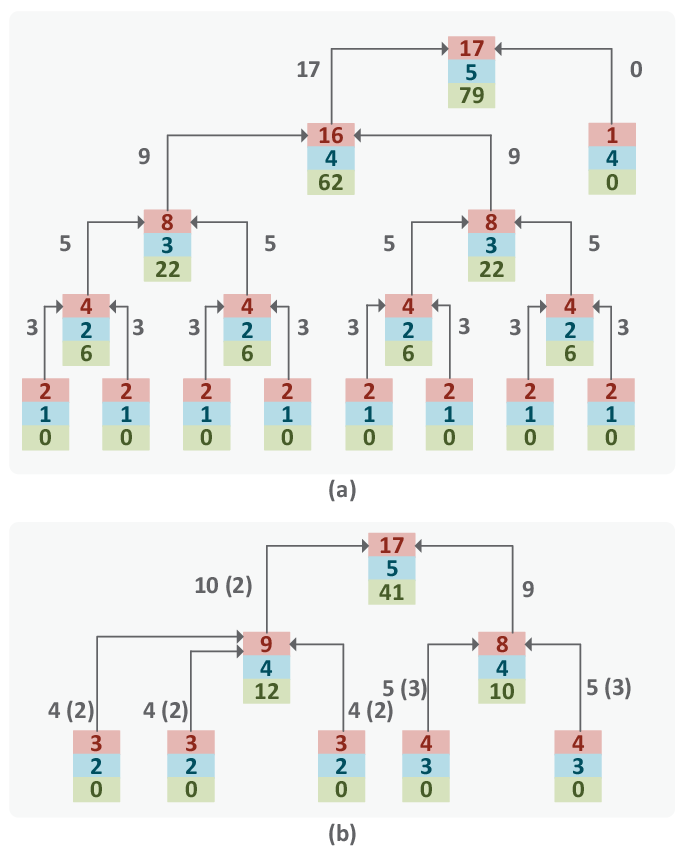}
        \vspace{-5pt}
        \caption{Tree structures for multiplying 17 ciphertexts: (a) a binary tree with the conventional single rescaling; (b) partitions with minimized number of rescaling incorporating multi-rescaling. The numbers in the red, blue, and green blocks of each node represent the number of ciphertext polynomials multiplied, the maximum multiplicative depth allowed, and the number of single rescaling units required at that node, respectively. For each edge, the number not in parenthesis indicates the number of rescaling units required. Among those, the numbers of combined rescaling units is listed in the parenthesis.}
        \vspace{-10pt}
\label{fig: tree structure for 17-PM}
\end{figure}

Consider the multiplication of 17 ciphertexts. The optimal multiplicative depth is $\lceil \log_2 17 \rceil = 5$. Figure~\ref{fig: tree structure for 17-PM}(a) depicts a binary tree structure for polynomial multiplication and rescaling using the single-rescaling approach described in \eqref{eq: 1 rs NTT}. In this figure, the numbers in the red, blue, and green blocks of each node represent the number of ciphertext polynomials multiplied, the maximum multiplicative depth allowed, and the number of single rescaling units required at that node, respectively. Each partition corresponds to a node connected to others via edges. For example, at the top of Fig.~\ref{fig: tree structure for 17-PM}(a), the 17 inputs are partitioned into (16, 1). The number on each edge indicates the number of rescaling units required. Take the edge from the node processing 16 ciphertext polynomials to the root node as an example. Since this node multiplies 16 ciphertexts, its output comprises 17 polynomials. Hence, 17 rescaling units are needed, one on each polynomial. Before this multiplication, 62 rescaling units are required for the lower layers of the tree. Besides, the other single input to the root of the tree requires no rescaling. Consequently, a total of $17 + 62 + 0 + 0 = 79$ rescaling units are needed. The 17+1=18 output polynomials of the root undergo relinearization, whose two output polynomials pass through two rescaling units to derive the final result.

Consider the partition shown in Fig.~\ref{fig: tree structure for 17-PM}(b). Its multiplicative depth is $\lceil \log_2 17\rceil = 5$. However, the nodes in the last layer of the tree on the left side handle three inputs each. Consequently, each corresponding polynomial multiplication is followed by two single rescalings, which are combined into a single 2-RS operation. In this figure, the digits in parentheses along the edges indicate the number of combined rescaling units, and the parentheses are omitted for edges involving only a single rescaling. Similarly, 3-RS blocks are applied after the multiplications of the two groups of four ciphertext
polynomials in the bottom right part of the tree, and a 2-RS is utilized after multiplying the three groups of polynomials to obtain the intermediate results of $9$ polynomials. Overall, the partition in Fig.~\ref{fig: tree structure for 17-PM}(b) only requires 41 rescaling units. As in part (a), the outputs of the root are passed through a relinearization block followed by two single rescalings to obtain the final result. Overall, the rescaling complexity for the 17-input ciphertext multiplication is reduced to $(41+2)/(79+2)=53\%$ by employing the optimized partition with multi-rescaling. 

\begin{table*}[t]
\centering
\caption{Comparisons on the numbers of rescaling ((I)NTT) operations in previous binary tree partition with single rescaling vs those in the proposed partition with the same multiplicative depth incorporating multi-rescaling.}
\label{tab: partitioning example}
\setlength{\tabcolsep}{4pt} % default is 6pt
\begin{tabular}{c||c|c|c|c|c}
\hline
\multirow{2}{*}{$n$} & \multirow{2}{*}{\begin{tabular}{@{}c@{}}
     Mult. \\
     Depth 
\end{tabular}} & \multicolumn{2}{c|}{Previous Binary Tree Partition} & \multicolumn{2}{c}{Proposed Optimized Partition} \\
\hhline{~|~|-|-|-|-|}
& & Partitions & \# of RS \big((I)NTTs\big) & Partitions & \# of RS \big((I)NTTs\big) \\ \hline\hline
$3$ & $2$ & $(2,1)$ & $3\+2$ ($3L+2L)$ & $(1,1,1)$ & $0+2$  ($0+2L$) \\ \hline
$4$ & $2$ & $(2,2)$ & $6\!\!+\!\!2$  ($6L\!\!+\!\!2(L\!\!-\!\!1)$) & $(2,2)$ & $6\!\!+\!\!2$  ($6L\!\!+\!\!2(L\!\!-\!\!1)$ )\\ \hline
$5$ & $3$ & $(4,1)\mid(2,2)$ & $11\!\!+\!\!2$  ($11L\!\!-\!\!5\!\!+\!\!2(L\!\!-\!\!2)$) & $(2,2,1)$ & $6\!\!+\!\!2$  ($6L\!\!+\!\!2(L\!\!-\!\!1)$) \\ \hline
$6$ & $3$ & $(4,2)\mid(2,2)$ & $14\!\!+\!\!2$ ($14L\!\!-\!\!8\!\!+\!\!2(L\!\!-\!\!2)$) & $(3,3)$ & $8\!\!+\!\!2$ ($8L\!\!+\!\!2(L\!\!-\!\!2)$) \\ \hline
$7$ & $3$ & $(4,3)\mid(2,2),(2,1)$ & $18\!\!+\!\!2$ ($18L\!\!-\!\!9\!\!+\!\!2(L\!\!-\!\!2)$) & $(4,3)\mid(2,2)$ & $15\!\!+\!\!2$ ($15L\!\!-\!\!5\!\!+\!\!2(L\!\!-\!\!2)$) \\ \hline
$8$ & $3$ & $(4,4)\mid(2,2),(2,2)$ & $22\!\!+\!\!2$ ($22L\!\!-\!\!10\!\!+\!\!2(L\!\!-\!\!2)$) & $(4,4)\mid(2,2),(2,2)$ & $22\!\!+\!\!2$ ($22L\!\!-\!\!10\!\!+\!\!2(L\!\!-\!\!2)$) \\ \hline
$9$ & $4$ & $(8,1)\mid(4,4)\mid(2,2),(2,2)$ & $31\!\!+\!\!2$ ($31L\!\!-\!\!28\!\!+\!\!2(L\!\!-\!\!3)$) & $(3,3,3)$ & $12\!\!+\!\!2$ ($12L\!\!+\!\!2(L\!\!-\!\!2)$) \\ \hline
$10$ & $4$ & $(8,2)\mid(4,4)\mid(2,2),(2,2)$ & $34\!\!+\!\!2$ ($34L\!\!-\!\!32\!\!+\!\!2(L\!\!-\!\!3)$) & $(3,3,4)\mid(2,2)$ & $19\!\!+\!\!2$ ($19L\!\!-\!\!5\!\!+\!\!2(L\!\!-\!\!2)$) \\ \hline
$11$ & $4$ & $(8,3)\mid(4,4),(2,1)\mid(2,2),(2,2)$ & $38\!\!+\!\!2$ ($38L\!\!-\!\!39\!\!+\!\!2(L\!\!-\!\!3)$) & $(4,4,3)\mid(2,2), (2,2)$ & $26\!\!+\!\!2$ ($26L\!\!-\!\!14\!\!+\!\!2(L\!\!-\!\!2)$) \\ \hline
$12$ & $4$ & $(8,4)\mid(4,4),(2,2)\mid(2,2),(2,2)$ & $42\!\!+\!\!2$ ($42L\!\!-\!\!44\!\!+\!\!2(L\!\!-\!\!3)$)   &$(6,6)\mid(3,3), (3,3)$ & $30\!\!+\!\!2$ ($30L\!\!-\!\!24\!\!+\!\!2(L\!\!-\!\!2)$) \\ \hline
\end{tabular}
\vspace{-10pt}
\end{table*}

The inputs can be partitioned in multiple ways. Various partition methods must be explored to identify the one with the lowest complexity, considering that several concatenated rescaling operations can be implemented with nearly the same complexity as a single rescaling, based on the combined scheme proposed in the previous subsection. The partition in Fig. \ref{fig: tree structure for 17-PM} is the one with the lowest complexity for 17 inputs. The number of inputs needed for practical applications is small or moderate. For input numbers ranging from 3 to 12, Table \ref{tab: partitioning example} compares the original partitioning strategy, which relies solely on single rescaling operations, with the proposed strategy that employs the multi-rescaling method of Algorithm \ref{alg: multi-RS} to minimize the total number of rescaling operations. In this table, partitions across different layers are separated by `$|$', while partitions within the same layer are separated by `$,$'. As an example, for $n=7$, the seven inputs are first partitioned into groups of $(4,3)$ in the first layer. In the second layer, the 4 and 3-input groups are further divided into $(2,2)$ and $(2,1)$, respectively; hence, the partition is denoted by $(4,3)|(2,2),(2,1)$. In the proposed design, since combined rescaling is utilized, the division can terminate at a larger group size. For $n=7$, the optimal partition is expressed as $(4,3)|(2,2)$. It means that although the 4-input is further partitioned into $(2,2)$ in the second layer, the 3-input group from the first layer is not further divided. The three inputs are multiplied altogether, and the outputs are subsequently processed through a two-level rescaling operation. Overall, the proposed partition tries to minimize the number of rescaling operations without increasing the multiplicative depth. In contrast to the original multi-input multiplication scheme that exclusively employs a binary tree, the optimal design from the proposed method may have a binary, ternary, or mixed tree structure.

It can be observed from Table \ref{tab: partitioning example} that the proposed partitioning method incorporating multi-rescaling achieves substantial complexity reduction, except for cases where $n$ is a power of two. When $n=2^a\ (a\in \mathbb Z^+)$, the inputs can be organized into a perfectly balanced binary tree. On the other hand, when $n$ is not a power of 2, as in the case of $n=17$ shown in Fig. \ref{fig: tree structure for 17-PM}, the tree becomes imbalanced in depth. This imbalance increases the multiplicative depth in certain computation branches and enables rescaling combination. The greater the imbalance of the tree, the more rescaling operations can be combined. Therefore, for $2^a<n<2^{a+1}$, the closer $n$ is to $2^a$, the greater complexity reduction can be achieved in terms of percentage.

In the preliminary version of this work \cite{3-ct_mult}, it was shown that the three-input multiplier reduces the upper bound of the noise in the product. The reduction is due to the fact that the noises associated with the evaluation keys are added, rather than multiplied, as in the conventional approach based on two-input multipliers. The improvements for three-input multiplier presented in Section \ref{sec: low-complexity ciphertext multiplication} do not compromise the noise level compared to that in \cite{3-ct_mult}. Similarly, for the multi-input ciphertext multiplication, the upper bound on the noise is also lower than that of the original method because the noises of the evaluation keys are added. The mathematical derivations are very similar to those in \cite{3-ct_mult} and are omitted here.

\section{Complexity Analyses and Comparisons} \label{sec: complexity analyses and comparisons}
This section first analyzes the complexities of the building components of our proposed architectures. Then the proposed designs are compared with our preliminary design for three inputs, as well as previous approaches for multiplying $n$ ciphertexts for $n$ ranging from 2 to 12.

\begin{table}[t]
\begin{center}\caption{Complexities of 2-parallel building blocks in ciphertext multiplications with $L$ and $K$ $w$-bit factors in $Q$ and $P$, respectively, and ring dimension $N$.}
\label{tab: computation components complexities}
\begin{tabular}{@{}c|@{}c@{}|@{}c@{}|@{}c@{}|@{}c@{}||@{}c@{}}
\hhline{~-----}
 &
\begin{tabular}{@{}c@{}} 
    Mod.\\
    Mult.
\end{tabular} 
 &
\begin{tabular}{@{}c@{}} 
    Mod. 
    \\ Adder
\end{tabular} &
\begin{tabular}{@{}c@{}} 
    Memory \\
    ($\times\,\,w$)
\end{tabular}
  & 
\begin{tabular}{@{}c@{}} 
    Registers \\
    ($w$-bit)
\end{tabular}
  &
\begin{tabular}{@{}c@{}} 
    Pipe. \\
    stages
\end{tabular} \\ \hline
NTT & $\log_2 N$ & $2\log_2 N$  & $2N-2$ 
&  
\begin{tabular}{@{}c@{}} 
    % $N-2+$ \\
    $10\!\log_2\!\!N$
\end{tabular}
& 
\begin{tabular}{@{}c@{}} 
    $N/2\!\!-\!\!1$ \\
    $+5\!\log_2\!\!N$
\end{tabular}\\ \hline
INTT & $\log_2 N$ & $4\log_2 N$ & $2N-2$ 
&  
\begin{tabular}{@{}c@{}} 
    % $N-2+$ \\
    $10\!\log_2\!\!N$
\end{tabular}
&
\begin{tabular}{@{}c@{}} 
    $N/2\!\!-\!\!1$ \\
    $+5\!\log_2\!\!N$
\end{tabular}\\ \hline
% BCove & $2L\!\!+\!\!2LK$ & $2K\!(\!L\!\!-\!\!1\!)$ & $L\!\!+\!\!LK$ 
% & 
% \begin{tabular}{@{}c@{}} 
%     $6L\!\!+\!\!K$ \\
%     $+\!6LK$
% \end{tabular}
% & $7$ \\ \hline
(sc)BConv & $2L\!\!+\!\!2LK$ & $2K\!(\!L\!\!-\!\!1\!)$ & $L\!\!+\!\!LK$ 
& 
\begin{tabular}{@{}c@{}} 
    $6L\!\!+\!\!K$ \\
    $+\!6LK$
\end{tabular}
& $7$\\ \hline
%CWPM & $2$ & 0 & 0 & $6$ & $3$ \\ \hline
%Cons. mult. & $2$ & 0 & 0 & $6$ & $3$ \\ \hline
\end{tabular}
\end{center}
\vspace{-10pt}
\end{table}

\subsection{Complexities of Building Blocks}\label{subsec: micro benchmark}

In our design, NTT is applied to polynomials, so their multiplications become coefficient-wise multiplications. (I)NTTs are also utilized in relinearization and rescaling. The 2-parallel NTT and INTT architectures of \parencite{ParhiNTT} are adopted in our design. Each NTT or INTT block consists of $\ log_2N$ processing elements (PEs) for processing one butterfly operation in each state. The PE in stage $s$ is connected to a memory unit of $2^s \times w$ to read the twiddle factors, where $w$ is the bit length of the modulus. There are also delay elements and multiplexers among the PEs to coordinate the data flow of the computations. Because the data of delay elements are accessed sequentially, these register chains used among the PEs in the (I)NTT design of \cite{ParhiNTT} are replaced with cyclic SRAM-based memory arrays in this paper. Each PE in the NTT contains one modular multiplier and two modular adders, whereas each of those in INTT includes one modular multiplier, four modular adders, and two multiplexers. Each modular multiplier is implemented using Barrett reduction with three pipelining stages \parencite{ParhiNTT}, resulting in a single $w$-bit multiplier in the critical path. Consequently, each PE in the NTT/INTT is pipelined into five stages. The complexities of the 2-parallel NTT, INTT, and (sc)BConv units are listed in Table~\ref{tab: computation components complexities}.

Based on Barrett reduction, a $w$-bit modular multiplier can be implemented by three $w$-bit multipliers, three modular adders, and one multiplexer \parencite{ModMulBarrett}. A modular adder can be realized using two adders, one comparator, and one multiplexer. A $w$-bit carry-ripple adder is built from $w$ full adders (FAs), while a $w$-bit multiplier requires $w(w-1)$ FAs. Each FA occupies an area equivalent to 4.5 XOR gates. The area of a multiplexer is assumed to be equal to that of one XOR gate. Similarly, a single-bit memory cell is considered equivalent in area to one XOR gate, whereas a delay element corresponds to the area of three XOR gates. These assumptions are used to estimate the gate counts of designs.

\begin{table}[t]
\begin{center}
\caption{Complexities of the proposed combined $\mu$-rescaling compared with $\mu$ single rescaling operations, assuming the number of moduli in $Q$ is $L$.}
\label{tab: multi-rs complexity}
\begin{tabular}{@{}c@{}|@{}c@{}|@{}c@{}|@{}c@{}}
\hhline{~---}
 &
\begin{tabular}{@{}c@{}} 
   \# of NTT
\end{tabular} 
 &
\begin{tabular}{@{}c@{}} 
   \# of INTT
\end{tabular} 
&
\begin{tabular}{@{}c@{}} 
   \# of Cons. mult.
\end{tabular}\\ \hline
\begin{tabular}{@{}c@{}} 
    $\mu$ single  \\
    RS
\end{tabular}
& 
\begin{tabular}{@{}c@{}} 
    $L-\mu$
\end{tabular}
& 
\begin{tabular}{@{}c@{}} 
    $L$
\end{tabular}
& 
\begin{tabular}{@{}c@{}} 
    $\mu L - \mu(\mu+1)/2$
\end{tabular} \\ \hline
\begin{tabular}{@{}c@{}} 
    Proposed\\
    combined $\mu$-RS
\end{tabular}
& 
\begin{tabular}{@{}c@{}} 
    $L-\mu$
\end{tabular}
& 
\begin{tabular}{@{}c@{}} 
    $\mu$
\end{tabular}
& 
\begin{tabular}{@{}c@{}} 
    $(\mu+1)L-\mu(\mu+3)/2$
\end{tabular}\\ \hline
\end{tabular}
\end{center}
\vspace{-10pt}
\end{table} 

Table~\ref{tab: multi-rs complexity} summarizes the complexities of the proposed Algorithm~\ref{alg: multi-RS} for combining $\mu$ stages of rescaling and compares them with those of the baseline implementation, which performs an INTT at the beginning, applies $\mu$ stages of rescaling according to \eqref{eq: 1 rs NTT}, and performs an NTT at the end. The proposed algorithm reduces the number of (I)NTTs by $L-\mu$ while introducing $L-\mu$ additional constant multiplications. Apparently, an (I)NTT is much more complicated than a polynomial multiplied by a constant. Therefore, the proposed design achieves substantial complexity reduction.

\begin{table*}[t]
\centering
\caption{Complexities of the improved three-input ciphertext multiplication compared with prior work \cite{3-ct_mult}, for $L$ and $K$ $w$-bit factors in $Q$ and $P$, respectively, and the ring dimension $N$.}
\label{tab: improved 3-ct mult}
\setlength{\tabcolsep}{4.5pt} % default is 6pt
\begin{tabular}{c||c|c|c|c||c|c|c|c}
\hhline{~--------}
 & \multicolumn{4}{c||}{Prior 3-input ciphertext multiplication \cite{3-ct_mult}} & \multicolumn{4}{c}{Improved 3-input ciphertext multiplication} \\
\hhline{~--------}
& PM & Relinearization & Rescaling & Total & PM & Relinearization & Rescaling & Total  \\ \hline\hline
NTT & 0 & $2L+2K$ & $4L-6$ & $6L+2K-6$ & 0 & $2K$ & $2L-4$ & $2L+2K-4$\\ \hline
INTT & 0 & $4L+2K$ & $4$ & $4L+2K+4$ & 0 & $4L+2K$ & $0$ & $4L+2K$ \\ \hline
sc(BConv) & 0 & $4$ & 0 & $4$ & 0 & $4$ & 0 & $4$ \\ \hline
Mod. Mult. & $16L$ & $12L+8K$ & $8L-12$ & $36L\!+\!8K\!-\!12$ & $16L$ & $8L+8K$ & $8L-12$ & $32L\!+\!8K\!-\!12$ \\ \hline\hline
\begin{tabular}{c}
     Latency  \\
    (\# of clks)
\end{tabular}  & $8$ & 
\begin{tabular}{c}
     $2N+22$  \\
     $+20\log_2N$
\end{tabular} 
&
\begin{tabular}{c}
     $2N+8$  \\
     $+20\log_2N$
\end{tabular} 
& 
\begin{tabular}{c}
     $4N+38$  \\
     $+40\log_2N$
\end{tabular} 
& $8$ &  
\begin{tabular}{c}
     $1.5N+18$  \\
     $+15\log_2N$
\end{tabular}
&
\begin{tabular}{c}
     $0.5N+8$  \\
     $+5\log_2N$
\end{tabular}
& 
\begin{tabular}{c}
     $2N+34$  \\
     $+20\log_2N$
\end{tabular}\\ \hline
\end{tabular}
\vspace{-5pt}
\end{table*}

\subsection{Improved three-input ciphertext multiplication}
This subsection compares the complexity of the proposed improved three-input ciphertext multiplication in Fig.~\ref{fig: 3 ct mult arc proposed low comp} with that of the prior work in \cite{3-ct_mult}, shown in Fig.~\ref{fig: 3 ct mult arc}. For the improved design in Fig. ~\ref{fig: 3 ct mult arc proposed low comp}, the ModUp block is implemented by the architecture in Fig.~\ref{fig: 2 ct mult arc}(b). The scBConv units is implemented according to \eqref{eq: scBconv} scaled by $P^{-1}\mod {q_j}$. 
Utilizing the 2-parallel fully pipelined (I)NTT designs from \cite{ParhiNTT}, a fully-pipelined 2-parallel architecture is achieved for the ciphertext multiplication in the RNS-CKKS scheme.

The hardware complexity of the improved three-input ciphertext multiplication algorithm, compared with that of the prior work in \cite{3-ct_mult}, is summarized in Table~\ref{tab: improved 3-ct mult}. The proposed design reduces the number of (I)NTT operations in the relinearization and rescaling steps by $2L$ and $2L + 2$, respectively. Overall, the total number of (I)NTTs is reduced by $4L + 2$. Since the (I)NTT blocks dominate the overall complexity of the ciphertext multiplier, the proposed design achieves a substantial reduction in area. Furthermore, one and three (I)NTT blocks are removed from the data paths of the relinearization and rescaling, respectively, reducing the number of (I)NTT operations in the data path of \cite{3-ct_mult} from 8 to 4. This shortens the data path and reduces the latency by approximately half. For both designs, memories are needed to store the evaluation keys, the three input ciphertexts, the twiddle factors needed for the (I)NTT operations, and the intermediate data exchanged among the PEs inside the (I)NTT units. Although the numbers of input ciphertexts and evaluation keys are identical in both designs, the proposed architecture reduces the number of (I)NTT operations, thereby lowering the overall memory requirement.

\begin{table}[t]
\centering
\caption{Synthesis results of three-input ciphertext multipliers for $L=K=24$, $N=2^{16}$, and $w=64$ using GlobalFoundries 22FDX process under the timing constraint of $2.5\text{ ns}$.}
\label{tab: improved 3-ct mult synth}
\setlength{\tabcolsep}{4.5pt} % default is 6pt
\begin{tabular}{c||c|c}
\hhline{~--}
& Prior work \cite{3-ct_mult} & 
Proposed work \\ \hline
PM ($\mathbf{mm}^2$) & 13.4 & 13.4 \\ \hline
 Relinearization ($\mathbf{mm}^2$) & 320.3 & 289.5 \\ \hline
 Rescaling ($\mathbf{mm}^2$) & 60.0 & 31.4 \\ \hline\hline
 Total logic area ($\mathbf{mm}^2$) & 393.7 & 334.3 \\ \hline\hline
 Latency (ms) & 0.66 & 0.33 \\ \hline\hline
 Memory (MB) & 404 & 502 \\ \hline 
\end{tabular}
\vspace{-5pt}
\end{table}

For further evaluation of the complexity, both the proposed 3-input ciphertext multiplier and that from \cite{3-ct_mult} are implemented in Verilog for $L=K=24$, $N=2^{16}$, and $w=64$. The designs are synthesized using the GlobalFoundries 22FDX process with GENUS 21.11, and the results are listed in Table \ref{tab: improved 3-ct mult synth}. Both designs have one 64x64-bit multiplier in the critical path, and the achievable tightest timing constraint is $2.5$ ns. It can be observed from Table \ref{tab: improved 3-ct mult synth} that the proposed multiplier in Fig. \ref{fig: 3 ct mult arc proposed low comp} achieves 15\% logic area saving and 50\% shorter latency compared to the architecture in Fig. \ref{fig: 3 ct mult arc} from \cite{3-ct_mult}. For both designs, two evaluation keys need to be stored, and they require $2\times 2\times (L+K)\times N\times w\approx$ 96 MB memory. Besides, the buffer for storing the three input ciphertexts is $3\times 2\times L\times N\times w\approx$ 72 MB. From Table \ref{tab: computation components complexities}, each (I)NTT needs $(2N-2)\times w\approx$ 1 MB, while the memory requirement of the sc(BConv) is relatively negligible. From the required number of (I)NTT units listed in Table \ref{tab: improved 3-ct mult}, it can be calculated that the previous design in Fig. \ref{fig: 3 ct mult arc} needs around 502 MB of memory. The proposed 3-input multiplier in Fig. \ref{fig: 3 ct mult arc proposed low comp} reduces the number of (I)NTT operations by $98$ according to Table \ref{tab: improved 3-ct mult}. As a result, it achieves a substantial reduction of $98/502\approx20\%$ in the memory requirement.

\subsection{Multi-input ciphertext multiplication}

\begin{table*}[t]
\begin{center}
\caption{Complexity of the proposed multi-input ciphertext multiplier for $L=K=24$, $N=2^{16}$, and $w=64$ compared to that of previous design built using a binary tree with 2-input ciphertext multipliers.}
\label{tab: n-ct mult}
\setlength{\tabcolsep}{4pt}
\begin{tabular}{c|c|c|c|c|c|c|c|c|c|c}
\hhline{~|----------|}
 & n & 4 & 5 & 6 & 7 & 8 & 9 & 10 & 11 & 12 \\ \hline
 \multirow{7}{*}{Baseline} & Memory (MB) & 525 & 692 & 859 & 1026 & 1193 & 1360 & 1527 & 1694 & 1861 \\ \cline{2-11}
 & 
 \begin{tabular}{@{}c@{}}
      Logic Area  \\
      (\# of XORs)($\times10^{10}$) 
 \end{tabular}
  & 0.838 & 1.117 & 1.396 & 1.676 & 1.955 & 2.234 & 2.514 & 2.793 & 3.072 \\ \cline{2-11}\noalign{\vskip 2pt} \cline{2-11}
 & 
 \begin{tabular}{@{}c@{}}
       Total Area  \\
      (\# of XORs)($\times10^{10}$) 
 \end{tabular} 
  & 1.278 & 1.698 & 2.117 & 2.536 & 2.956 & 3.375 & 3.795 & 4.214 & 4.633 \\ \cline{2-11}
 & 
 \begin{tabular}{@{}c@{}}
       Latency \\
      (\# of clks) 
 \end{tabular} 
  & 262830 & 394245 & 394245 & 394245 & 394245 & 525660 & 525660 & 525660 & 525660 \\ \hline\hline
 \multirow{7}{*}{Proposed} & Memory (MB) 
 &
 \begin{tabular}{@{}c@{}}
       455 \\
      (86.7\%) 
 \end{tabular}
 &
    \begin{tabular}{@{}c@{}}
       550 \\
      (79.5\%) 
 \end{tabular} 
 &
 \begin{tabular}{@{}c@{}}
       671 \\
      (78.1\%) 
 \end{tabular} 
 &
 \begin{tabular}{@{}c@{}}
       849 \\
      (82.7\%) 
 \end{tabular} 
 &
 \begin{tabular}{@{}c@{}}
       1026 \\
      (86.0\%) 
 \end{tabular} 
 & 
 \begin{tabular}{@{}c@{}}
       1006 \\
      (74.0\%) 
 \end{tabular} 
 &
 \begin{tabular}{@{}c@{}}
       1184 \\
      (77.5\%) 
 \end{tabular} 
 &
 \begin{tabular}{@{}c@{}}
       1361 \\
      (80.3\%) 
 \end{tabular} 
 &
 \begin{tabular}{@{}c@{}}
       1497 \\
      (80.4\%) 
 \end{tabular}\\ \cline{2-11}
 & 
 \begin{tabular}{@{}c@{}}
      Logic Area  \\
      (\# of XORs)($\times10^{10}$) 
 \end{tabular} 
 & 
 \begin{tabular}{@{}c@{}}
       0.611 \\
      (72.9\%) 
 \end{tabular} 
 & 
 \begin{tabular}{@{}c@{}}
       0.664 \\
      (59.4\%) 
 \end{tabular} 
 & 
 \begin{tabular}{@{}c@{}}
       0.804 \\
      (57.6\%) 
 \end{tabular} 
 & 
 \begin{tabular}{@{}c@{}}
       1.097 \\
      (65.5\%) 
 \end{tabular}
 & 
 \begin{tabular}{@{}c@{}}
       1.390 \\
      (71.1\%) 
 \end{tabular}
 & 
 \begin{tabular}{@{}c@{}}
       1.145 \\
      (51.2\%) 
 \end{tabular} 
 & 
 \begin{tabular}{@{}c@{}}
       1.437 \\
      (57.2\%) 
 \end{tabular} 
 & 
 \begin{tabular}{@{}c@{}}
       1.730 \\
      (62.0\%) 
 \end{tabular} 
 & 
 \begin{tabular}{@{}c@{}}
       1.883 \\
      (61.3\%) 
 \end{tabular}  \\ \cline{2-11}\noalign{\vskip 2pt}\cline{2-11}
 & 
 \begin{tabular}{@{}c@{}}
       Total Area  \\
      (\# of XORs)($\times10^{10}$) 
 \end{tabular}
 & 
 \begin{tabular}{@{}c@{}}
       0.992 \\
      (77.6\%) 
 \end{tabular} 
 & 
 \begin{tabular}{@{}c@{}}
       1.125 \\
      (66.3\%) 
 \end{tabular} 
 & 
 \begin{tabular}{@{}c@{}}
       1.367 \\
      (64.6\%) 
 \end{tabular} & 
 \begin{tabular}{@{}c@{}}
       1.809 \\
      (71.3\%) 
 \end{tabular}
 & 
 \begin{tabular}{@{}c@{}}
       2.251 \\
      (76.1\%) 
 \end{tabular}
 & 
 \begin{tabular}{@{}c@{}}
       1.989 \\
      (58.9\%) 
 \end{tabular} & 
 \begin{tabular}{@{}c@{}}
       2.430 \\
      (64.0\%) 
 \end{tabular} & 
 \begin{tabular}{@{}c@{}}
       2.872 \\
      (68.2\%) 
 \end{tabular} & 
 \begin{tabular}{@{}c@{}}
       3.138 \\
      (67.7\%) 
 \end{tabular} \\ \cline{2-11}
 & 
 \begin{tabular}{@{}c@{}}
       Latency  \\
      (\# of clks) 
 \end{tabular} 
 & 
 \begin{tabular}{@{}c@{}}
       197119 \\
      (75.0\%) 
 \end{tabular} 
 & 
 \begin{tabular}{@{}c@{}}
       197120 \\
      (50.0\%) 
 \end{tabular} 
 & 
 \begin{tabular}{@{}c@{}}
       197125 \\
      (50.0\%) 
 \end{tabular} & 
 \begin{tabular}{@{}c@{}}
       262828 \\
      (66.7\%) 
 \end{tabular}
 & 
 \begin{tabular}{@{}c@{}}
       262825 \\
      (66.7\%) 
 \end{tabular}
 & 
 \begin{tabular}{@{}c@{}}
       197136 \\
      (37.5\%) 
 \end{tabular} 
 & 
 \begin{tabular}{@{}c@{}}
       262835 \\
      (50.0\%) 
 \end{tabular} 
 & 
 \begin{tabular}{@{}c@{}}
       262836 \\
      (50.0\%) 
 \end{tabular} 
 & 
 \begin{tabular}{@{}c@{}}
       262833 \\
      (50.0\%) 
 \end{tabular} \\ \hline
 % \hline
 % & Normalized ADP & 61.9\% & 35.6\% & 34.8\% & 51.5\% & 55.3\% & 24.0\% & 35.0\% & 37.3\% & 37.1\% \\ \cline{2-11}
\end{tabular}
\end{center}
\end{table*}

In this subsection, the complexity of the proposed multi-input ciphertext multiplier utilizing the optimized relinearization and combined rescaling is analyzed for $n$ ranging from 4 to 12 with parameters $L=K=24$, $N=2^{16}$, and $w=64$. The optimized partitions in Table \ref{tab: partitioning example} are employed, and the resulting complexities are reported in Table \ref{tab: n-ct mult}. The area is estimated in terms of the number of XOR gates, and the total area is the sum of the logic and memory areas. Our design is fully pipelined with up to one $w$-bit multiplier in the critical path. The latencies are listed in terms of clock cycles. 

As no prior multi-input ciphertext multiplier exists, our proposed design is compared with a multi-input multiplier built using a binary tree of two-input ciphertext multipliers, as summarized in Table~\ref{tab: computation components complexities}. This multiplier requires $n - 1$ two-input ciphertext multipliers, each comprising polynomial multiplication, relinearization, and rescaling. Therefore, its overall complexity increases linearly with $n$, and its latency equals $\lceil \log_2 n \rceil$ times that of a two-input multiplier.

For each design, memories are required to store the input ciphertexts, the twiddle factors for the (I)NTT operations, and the evaluation keys. As the number of inputs $n$ increases, the memory demand generally grows. However, in our proposed design, the memory requirement for $n=9$ is lower than that of $n=8$. Although the $n=9$ case uses more memory blocks to store inputs and evaluation keys, its $(3,3,3)$ partition enables more combined rescaling, thereby reducing the number of (I)NTTs. Analyses for $n=16$ and $n=17$ show a similar trend. Due to the imbalance of the larger binary tree when $n=2^i+1$ ($2<i\in \mathbb{Z}^+$), the proposed design can take more advantage of combined rescaling, resulting in lower memory usage compared to the case of $n=2^i$. Compared to a multi-input multiplier built with a binary tree of 2-input multipliers, the proposed design achieves approximately 14\% memory reduction or more. The memory saving is lower when $n=2^i$ because the binary tree is perfectly balanced, and combined rescaling cannot be applied. The logic complexity and latency exhibit the same trend: both are lower for $n=2^i+1$ than for $n=2^i$. Unlike memory, the logic complexity and latency do not include any components that remain the same or increase compared to the design constructed by the 2-input multipliers. Therefore, they are reduced by larger percentages in our design. On average, for $n$ ranging from 4 to 12, the proposed design achieves 32\% and 45\% reductions on the overall area and latency, respectively.

The achievable latency reduction of the proposed design remains about the same for different values of $L$, $K$, and $N$, since the number of (I)NTTs in the data path does not vary with these parameters. For smaller $L$ and $K$, the proposed design can achieve slightly more reductions in memory usage and logic area because some rescaling operations are combined while others are not. Accordingly, constants such as -6, +4, and -4 listed in Table \ref{tab: improved 3-ct mult} appear in the total number of (I)NTTs, and their effects become more pronounced for smaller $L$ and $K$. For smaller $N$, the proposed design also achieves slightly higher area savings, as the (I)NTTs dominate the overall complexity. As shown in Table \ref{tab: computation components complexities}, the complexity of an (I)NTT block scales linearly with $N$.

\section{Conclusion} \label{sec: conclusion}
This paper first proposes improvements to the three-input ciphertext multiplication for the RNS-CKKS scheme. Subsequently, a framework for multi-input ciphertext multiplication is developed. New evaluation keys are introduced, and the relinearization process is reformulated to share intermediate results, thereby reducing the complexity of modulus switching. In addition, an efficient combined rescaling algorithm is proposed to perform multiple rescalings with nearly the same complexity as a single one. Guidelines are provided for input partitioning to enable the adoption of the multi-rescaling approach. Compared with the best prior design, the proposed scheme achieves significant reductions in both area and latency. Future work will further simplify the multiplier with a larger number of inputs. 

\printbibliography

\begin{IEEEbiography}[{\includegraphics[width=1in,height=1.25in, clip,keepaspectratio]{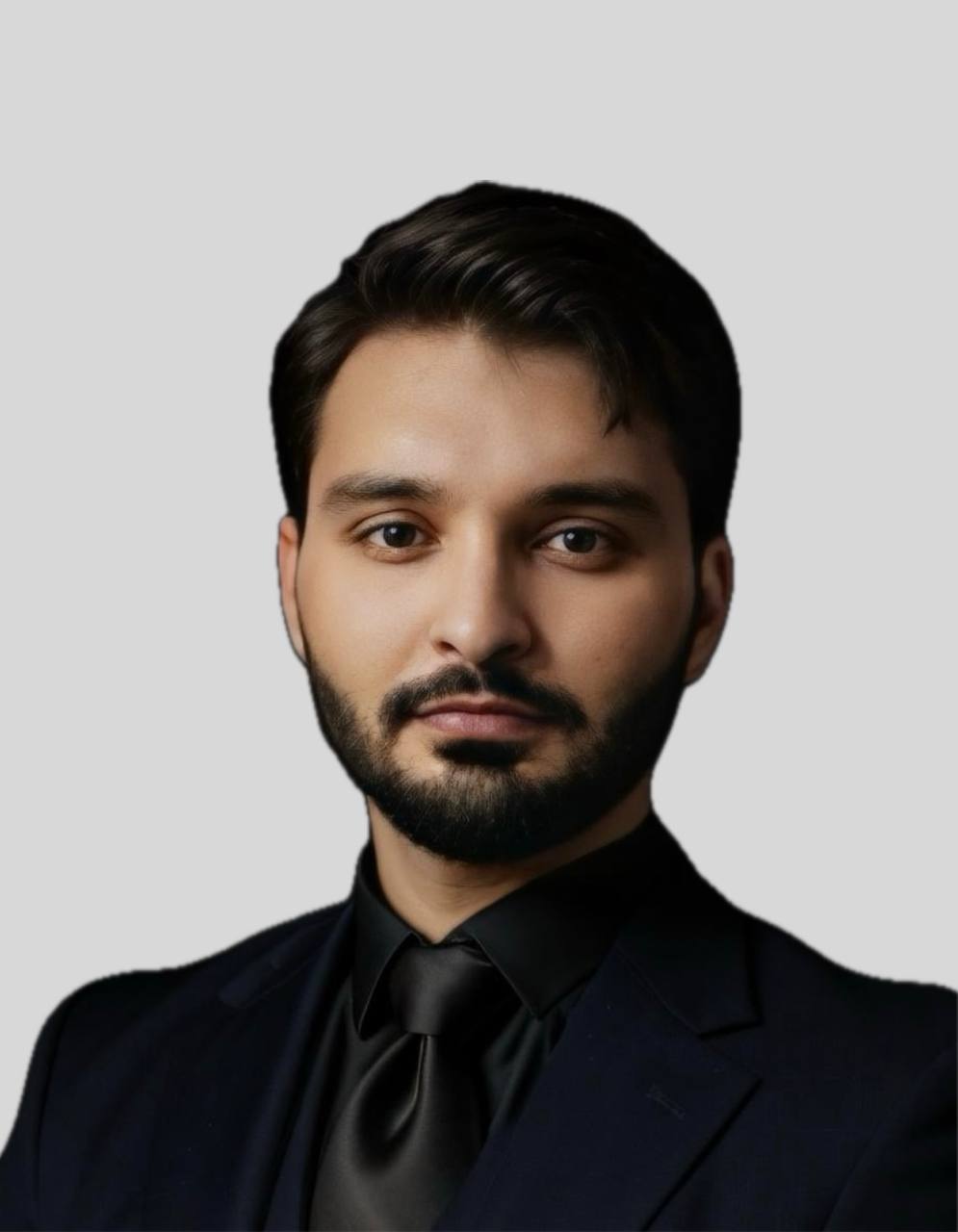}}]
{Sajjad Akherati} (Graduate Student Member, IEEE) received the B.Sc. degree in electrical engineering from Sharif University of Technology, Iran, in 2022. He is currently pursuing his Ph.D. degree with the Electrical and Computer Engineering Department, The Ohio State University, Columbus, OH, USA. His current research interests include VLSI architectures design for Homomorphic Encryption.
\end{IEEEbiography}
% \vspace{-400pt}
\begin{IEEEbiography}[{\includegraphics[width=1in,height=1.25in, clip,keepaspectratio]{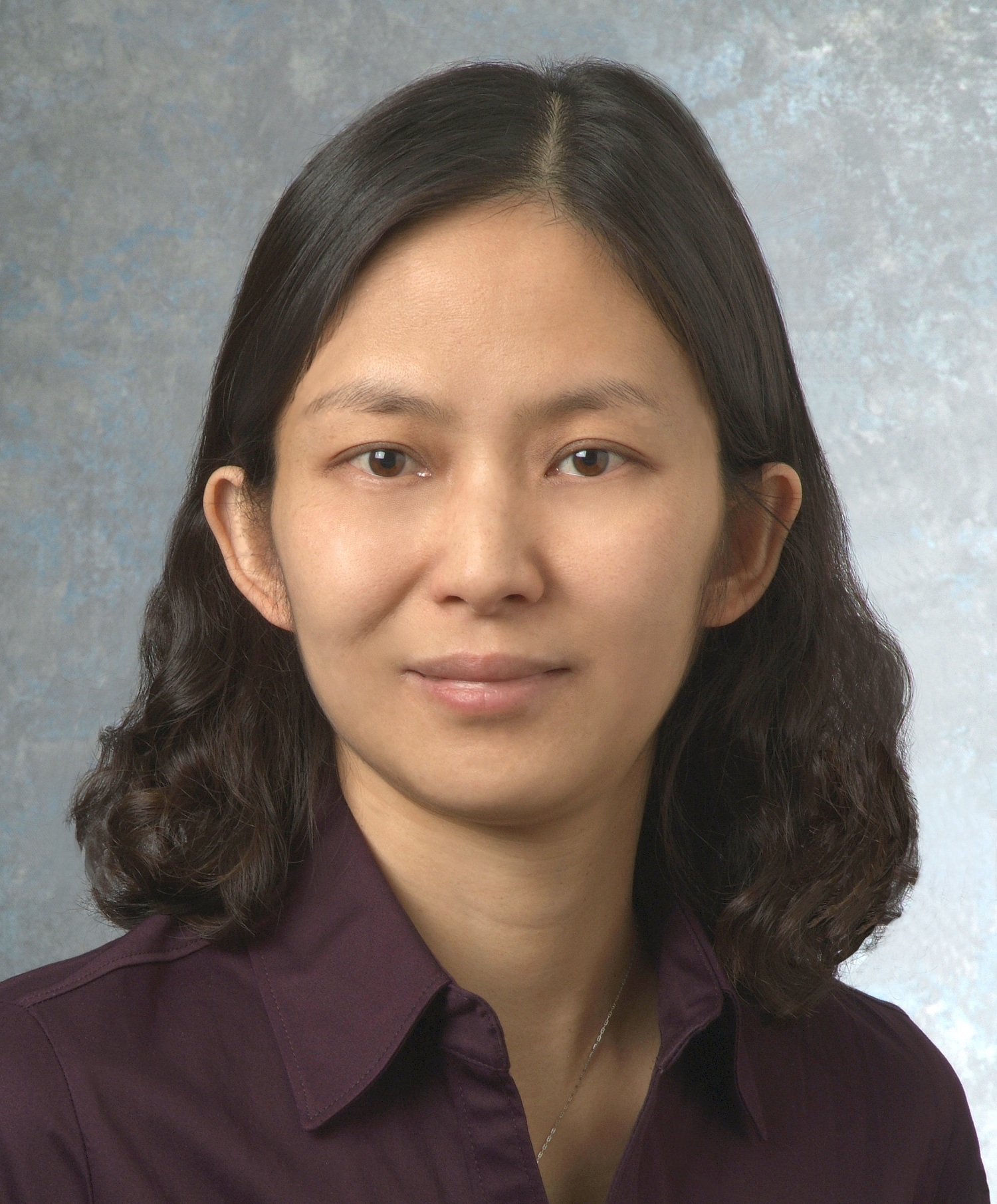}}]
{Xinmiao Zhang} (Fellow, IEEE) received her Ph.D. degree in Electrical Engineering from the University of Minnesota. She is currently a Professor at the Ohio State University. Prof. Zhang’s research spans the areas of VLSI architecture design, digital storage and communications, cryptography, machine learning, and signal processing. Prof. Zhang is a recipient of the NSF CAREER Award 2009, and the College of Engineering Lumley Research Award at The Ohio State University 2022. Prof. Zhang was elected the Vice President-Technical Activities of the IEEE Circuits and Systems Society (CASS) 2022-2023 and served on the Board of Governors of CASS 2019-2021. She also served on the technical program and organization committees of many conferences, including ISCAS, ICC, GLOBECOM, SiPS, GlobalSIP, MWSCAS, and GLSVLSI. She is the Associate Editor-in-Chief for the IEEE Transactions on Circuits and Systems-I 2024-2027.
\end{IEEEbiography}

\end{document}